\documentclass[prc,preprint,showpacs,preprintnumbers,showkeys,amsmath,amssymb]{revtex4}
\usepackage{graphicx}
\newcommand{\beq}{\begin{equation}}
\newcommand{\eeq}{\end{equation}}
\newcommand{\beqa}{\begin{eqnarray}}
\newcommand{\eeqa}{\end{eqnarray}}
\newcommand{\nn}{\nonumber}
\newcommand{\Sigs}{\Sigma_{\mathrm s} }
\newcommand{\Sigv}{\Sigma_{\mathrm v} }
\newcommand{\Sigo}{\Sigma_{\mathrm o} }
\newcommand{\kf}{k_{\mathrm F} }
\newcommand{\kfj}{k_{\mathrm Fj} }
\newcommand{\kfn}{k_{\mathrm Fn} }
\newcommand{\kfp}{k_{\mathrm Fp} }
\newcommand{\bfgamma}{\mbox{\boldmath$\gamma$\unboldmath}}
\newcommand{\veck}{\textbf{k}}
\newcommand{\vecp}{\textbf{p}}
\newcommand{\vecq}{\textbf{q}}
\newcommand{\vecu}{\textbf{u}}
\newcommand{\vecv}{\textbf{v}}

\newcommand{\pabs}{|{\bf p}|}
\newcommand{\qabs}{|{\bf q}|}
\newcommand{\Sigsn}{\Sigma_{{\mathrm s},n} }

\newcommand{\Sigon}{\Sigma_{{\mathrm o},n} }
\newcommand{\Sigsp}{\Sigma_{{\mathrm s},p} }

\newcommand{\Sigop}{\Sigma_{{\mathrm o},p} }
\newcommand{\Sigsi}{\Sigma_{{\mathrm s},i} }
\newcommand{\Sigvi}{\Sigma_{{\mathrm v},i} }
\newcommand{\Sigoi}{\Sigma_{{\mathrm o},i} }
\newcommand{\SigsDDRMFi}{\Sigma^{DDRMF}_{{\mathrm s},i} }
\newcommand{\SigoDDRMFi}{\Sigma^{DDRMF}_{{\mathrm 0},i} }
\begin{document}
\preprint{}
\title{Dirac-Brueckner-Hartree-Fock calculations for isospin asymmetric nuclear matter based on improved approximation schemes.}
\author{E. N. E. van Dalen}
\author{C. Fuchs}
\author{Amand Faessler}
\affiliation{Institut
f$\ddot{\textrm{u}}$r Theoretische Physik, Universit$\ddot{\textrm{a}}$t
T$\ddot{\textrm{u}}$bingen,
Auf der Morgenstelle 14, D-72076 T$\ddot{\textrm{u}}$bingen, Germany}
\begin{abstract}
We present Dirac-Brueckner-Hartree-Fock calculations for isospin asymmetric nuclear matter which are based on improved approximations schemes.
The potential matrix elements have been adapted for isospin asymmetric nuclear matter in order to account for the proton-neutron mass splitting in a more consistent way. The proton properties are particularly sensitive to this adaption and its consequences, whereas the neutron properties remains almost unaffected in neutron rich matter.  Although at present full Brueckner calculations are still too complex to apply to finite nuclei, these relativistic Brueckner results can be used as a guidance to construct a density dependent relativistic mean field theory, which can be applied to finite nuclei. It is found that an accurate reproduction of the Dirac-Brueckner-Hartree-Fock equation of state requires a renormalization of these coupling functions. 
\end{abstract}
\pacs{21.65.+f,21.60.-n,21.30.-x,21.30.Fe}
\keywords{Nuclear equation of state, isospin dependence, relativistic Brueckner approach,  density dependent relativistic mean field theory}
\maketitle
\section{Introduction}
The investigation of asymmetric
matter is of importance for astrophysical and nuclear structure studies.
In the field of astrophysics this investigation is important for the physics of supernova
explosions~\cite{bethe90} and of neutron
stars~\cite{pethick95}, e.g. the chemical composition
and cooling mechanism of protoneutron stars~\cite{lattimer91,vandalen06b},
mass-radius correlations~\cite{prakash88,zuo04}, and some other topics.
In the field of nuclear structure  the investigation of isospin asymmetric
matter is of interest in study of  neutron-rich
nuclei~\cite{tanihata95}.
This isovector dependence of the nuclear force can be investigated in the heavy ion experiments~\cite{baran05}.
However, the data for neutron-rich nuclei were rather scarce in the past. This situation is changing 
with the forthcoming new generation of radioactive
beam facilities, e.g. the future GSI facility FAIR in Germany, the Rare Isotope
Accelerator planned in the United States of America or SPIRAL2 at GANIL/France,
which will produce large amounts of new data.

Models which make predictions on the nuclear equation of state (EoS) can roughly be divided
into three classes: Phenomenological density functionals, effective field theory (EFT) approaches, and
ab initio approaches. Phenomenological density functionals are based on
effective density dependent interactions such as
Gogny or Skyrme forces \cite{reinhard04} or
relativistic mean field (RMF) models \cite{rmf} with usually more than six and less than 15
parameters.  The effective field theory approaches lead to a more systematic expansion of the
EoS in powers of density, respectively the Fermi momentum $k_F$. The advantage of EFT
is the small number of free parameters and a correspondingly higher
predictive power. 
Ab initio approaches are based on high precision
free space nucleon-nucleon interactions and the nuclear many-body
problem is treated microscopically. Predictions for the nuclear EoS
are essentially parameter free. Examples are variational
calculations \cite{akmal98}, Brueckner-Hartree-Fock (BHF)
\cite{lejeune00,zuo04}
or relativistic Dirac-Brueckner-Hartree-Fock (DBHF)
\cite{terhaar87,bm90,dejong98,gross99,alonso03,vandalen04b,vandalen05a,vandalen05b}
calculations and Greens functions
Monte-Carlo approaches \cite{carlson03}.

Many-body calculations,
on the other hand,
have to rely on the summation of relevant diagram classes and are still
too involved for systematic applications to finite nuclei. However, these results can be used as a guidance for the construction of a "semi"-phenomenological density functional. Examples are e.g. Gogny forces~\cite{gogny} derived
from $G$-matrices or density dependent relativistic mean field (DDRMF) theory~\cite{fuchs95,hofmann01}, which can be based on DBHF results.

The theoretical predictions for the isospin dependence of nuclear interactions are still diverse. 
RMF theory can not describe the complex nonlinear behavior of the DBHF and BHF binding energy at densities near $\rho=0$. Furthermore, the symmetry energy in relativistic DBHF calculations is found to be significantly stiffer than
in non-relativistic BHF approaches~\cite{bombaci91}, in particular at high densities.
The BHF calculations~\cite{muether02}
predict a proton-neutron mass splitting of $m^*_{NR,n} > m^*_{NR,p}$ in
neutron dominated  nuclear matter. In contrast, RMF theory with the scalar isovector $\delta$-meson included
predict the opposite behavior, $m^*_{D,n} < m^*_{D,p}$ \cite{baran05,liu02}.
The various Skyrme forces give opposite predictions
for the neutron-proton mass splitting. Relativistic {\it ab initio} calculations based
on realistic nucleon-nucleon interactions, as for instance the DBHF approach,
are the proper tool to answer these questions.

In this work we describe asymmetric nuclear  matter  in the framework of the
relativistic DBHF approach based on projection techniques using the Bonn
potential and their bare $NN$ matrix elements $V$~\cite{machleidt89}.
Furthermore, the optimal representation scheme for the $T$-matrix, the subtracted $T$
matrix representation, is applied. This scheme has previously been applied to asymmetric nuclear matter in refs.~\cite{vandalen04b,vandalen05a,vandalen05b}. However, in the present work we go beyond the approach used in~\cite{vandalen04b,vandalen05a,vandalen05b} in the sense that we improve at a couple of approximations. To be more precise, the Bonn potential has now been adapted for asymmetric nuclear matter. 

In the solution of Bethe-Salpeter (BS) equation we abandon the approximation of an averaged $np$ mass in the $np$ channel and distinguish explicitly between the different isospin dependent matrix elements. As a consequences, the potential and $T$-matrix are evaluated in terms of six independent helicity or covariant amplitudes instead of five~\cite{dejong98}, which are sufficient in the case of an averaged $np$ mass.

\indent The plan of this paper is as follows. The relativistic DBHF approach with emphasis on the treatment of the  $nn$, $pp$, and $np$ channels
is treated in sect.~\ref{sec:DBHF}. Results are presented in
sect.~\ref{sec:rd}. Furthermore, the relation between DBHF results and the RMF theory is discussed in sect.~\ref{sec:rrmft}. Finally, we end with a summary and a conclusion in
sect.~\ref{sec:c}.
\section{DBHF approach in isospin asymmetric nuclear matter}
\label{sec:DBHF}
In this section the relativistic Brueckner approach is discussed.
First a general overview is given, followed by a more detailed discussion of the modifications, which are necessary to account properly for the proton-neutron mass splitting and the isospin dependence of the corresponding matrix elements.

In the relativistic DBHF approach a nucleon inside
nuclear matter is regarded as a dressed particle
as a consequence of its interaction with the surrounding
nucleons. This interaction of the nucleons is treated in the ladder
approximation of the relativistic BS equation
\beqa
T = V + i \int  V Q G G T,
\label{subsec:SM;eq:BS}
\eeqa
where $T$ denotes the $T$-matrix, $V$ the bare nucleon-nucleon,  $Q$ the Pauli operator, and
$G$ the Green's function of an intermediate off-shell nucleon.
This Green's function $G$ which describes the propagation of dressed
nucleons in the medium fulfills the Dyson equation
\beqa
G=G_0+G_0\Sigma G,
\label{subsec:SM;eq:Dysoneq}
\eeqa
where $G_{0}$ denotes the free nucleon propagator and $\Sigma$ the self-energy.
In the Hartree-Fock approximation this self-energy is given by
\beqa
\Sigma = -i \int\limits_{F} (Tr[G T] - GT ).
\label{subsec:SM;eq:HFselfeq1}
\eeqa
The eqs.~(\ref{subsec:SM;eq:BS})-(\ref{subsec:SM;eq:HFselfeq1}) are strongly coupled. Therefore,
this set of equations represents a self-consistency problem and has to be iterated until
convergence is reached.

The structure of the self-energy follows from the requirement of translational and rotational invariance,
hermiticity, parity conservation, and time reversal invariance.
The most general form of the Lorentz structure of the self-energy  in
the nuclear matter rest frame is given by
\beqa
\Sigma(k,\kf)= \Sigs (k,\kf) -\gamma_0 \, \Sigo (k,\kf) +
\bfgamma  \cdot \textbf{k} \,\Sigv (k,\kf),
\label{subsec:SM;eq:self1}
\eeqa
where the $\Sigs$, $\Sigo$, and $\Sigv$ components are Lorentz scalar
functions which depend on the Lorentz invariants $k^2$,$k
\cdot  j$ and $j^2$, with $j_{\mu}$ the baryon
current. Therefore, these Lorentz invariants can be expressed in terms of
$k_0$, $|\veck|$ and $\kf$, where $\kf$ denotes the Fermi momentum.
The different components of the self-energy are determined by taking the
respective traces \cite{horowitz87,sehn97}
\beqa
\Sigs = \frac{1}{4} tr \left[ \Sigma \right],\quad
\Sigo = \frac{-1}{4} tr \left[ \gamma_0 \, \Sigma \right], \quad
\Sigv =  \frac{-1}{4|\veck|^2 }
tr \left[{\bfgamma}\cdot \veck \, \Sigma \right].
\label{subsec:SM;eq:trace}
\eeqa
\\ \indent The presence of the medium influences the masses and momenta of the nucleons inside nuclear matter.
These effective masses and effective
momenta of the nucleons can be written as
\beqa
m^*(k, \kf) = M + \Re e \Sigma_s(k, \kf), \quad k^{*\mu}=k^{\mu} + \Re e
\Sigma^{\mu}(k, \kf).
\label{subsec:SM;eq:dirac}
\eeqa
By the introduction of reduced quantities, one has the reduced effective mass
\beqa
{\tilde m}^*(k,\kf) = m^*(k,\kf)/ \left( 1+\Sigv(k,\kf)\right).
\label{subsec:SM;eq:redquantity}
\eeqa
and the reduced kinetic momentum
\beqa
{{\tilde k}^*}_\mu = k^*_\mu / \left( 1+\Sigv(k,\kf)\right) \, , \,
\eeqa
Hence, The Dirac equation written in terms of these reduced effective masses and
momenta has the form
\beqa
 [ \gamma_\mu {\tilde k}^{*^\mu} - {\tilde m}^*(k,\kf)] u(k,\kf)=0.
\quad
\label{subsec:SM;eq:dirac2}
\eeqa
To simplify  the self-consistency scheme we will work in the quasi-particle approximation,
i.e. the imaginary part of the self-energy $\Im m \Sigma$ will be
neglected.
In addition, the ``reference spectrum approximation''~\cite{bethe63} is applied
in the iteration procedure, i.e. the effective
mass of the nucleon is assumed to be entirely density dependent
($|\veck|=k_F$). However, in general the reduced effective mass is density and momentum
dependent. Therefore, this method implies that the self-energy
itself is only weakly momentum dependent. At the end of the calculation one has of course
to verify the consistency of the
assumption $\Sigma(k)\approx \Sigma(|\veck|=\kf)$
with the result of the iteration procedure.
\\ \indent The solution of the Dirac equation in eq.~(\ref{subsec:SM;eq:dirac2}) provides the
positive-energy in-medium nucleon spinor
\beqa
u_\lambda (k,\kf)= \sqrt{ { {\tilde E}^*(\veck)+ {\tilde m}^*_F}\over
{2{\tilde m}^*_F}}
\left(
\begin{array}{c} 1 \\
{2\lambda |\veck|}\over{{\tilde E}^*(\veck)+ {\tilde m}^*_F}
\end{array}
\right)
\chi_\lambda,
\label{spinor}
\eeqa
where ${\tilde E}^*(\veck)=\sqrt{\veck^2+{\tilde m}^{*2}_F}$ denotes the reduced effective energy
and $\chi_\lambda$ a two-component Pauli spinor with
$\lambda=\pm {1\over 2}$~\footnote{From now on we omit the tilde in this section because in the following we
  normally deal with ${\tilde m}^*_F,{\tilde k}^{*^\mu}$.}.
The normalization of the Dirac spinor is
thereby chosen as $\bar{u}_\lambda(k,\kf) u_\lambda(k,\kf)=1$.

\subsection{$nn$ and $pp$ channel}

It is convenient to reduce a four-dimensional BS integral equation, eq.~(\ref{subsec:SM;eq:BS}),
to a three-dimensional one to solve the
scattering problem of two nucleons in the nuclear medium. Therefore, the
two-particle propagator $iGG$ in the BS equation has to
be replaced by the effective Thompson propagator.
The Thompson propagator implies that the time-like component of the
momentum transfer in $V$ and $T$ is set equal to zero. Hence, the Thompson propagator
restricts the exchanged energy transfer by
$\delta(k^0)$ to zero. In addition, the Thompson propagator projects the intermediate nucleons onto
positive energy states. Thus, in the two-particle center of mass (c.m.)
frame, which is the natural frame for studying two-particle
scattering processes, the Thompson equation can be written as~\cite{terhaar87,sehn97}
\beqa
T(\vecp,\vecq,x)|_{c.m.} &=&  V(\vecp,\vecq)
\label{subsec:SM;eq:thompson} \\
&+&
\int {d^3\veck\over {(2\pi)^3}}
{\rm V}(\vecp,\veck)
{m^{*2}_F\over{E^{*2}(\veck)}}
{{Q(\veck,x)}\over{2{E}^*(\vecq)-2{E}^*(\veck)
+i\epsilon}}
T(\veck,\vecq,x),
\nonumber
\eeqa
where $\vecq=(\vecq_1 - \vecq_2)/2$ is the relative three-momentum
of the initial state and $\veck$ and $\vecp$ are the relative
three-momenta of the intermediate and the final states, respectively.
\\ \indent  The Thompson equation~(\ref{subsec:SM;eq:thompson})  can be solved applying
standard techniques, which are outlined in detail by Erkelenz~\cite{erkelenz74}.
To determine the self-energy only positive-energy states are
needed. Therefore, it is more convenient to apply the Dirac nucleon propagator~\cite{horowitz87},
\beqa
G_D(k, \kf)=[\gamma_{\mu} k^{* \mu} + m^*(k, \kf)] 2 \pi i \delta(k^{* 2} - m^{*
2}(k, \kf)) \Theta(k^{*0}) \Theta(k_F-|\veck|).,
\label{eq:G_D}
\eeqa
instead of the full nucleon propagator. Due to the $\Theta$-functions in eq.~(\ref{eq:G_D})
only positive energy nucleons are allowed in the intermediate scattering states.
In this way, one avoids the delicate problem of infinities in the
theory which generally will occur if one includes contributions
from negative energy nucleons in the Dirac sea~\cite{horowitz87,dejong98}.
\\ \indent In the on-shell case for identical particles only five of the sixteen helicity
matrix elements are independent which follows from general
symmetries~\cite{erkelenz74}.
After a partial wave projection onto the $|JMLS>$-states the Thomas equation
reduces to a set of one-dimensional integral equations over the relative momentum
$|\veck|$. Furthermore,  it decouples into three
subsystems of integral
equations: the uncoupled spin singlet, the uncoupled spin triplet,
and the coupled triplet states (appendix~\ref{app:pwd}). To achieve this reduction to the
one-dimensional integral equations the Pauli operator
$Q$ is replaced by an angle-averaged Pauli operator
$\overline{Q}$ ~\cite{horowitz87}. Due to  deformation of the Fermi sphere to a
Fermi ellipsoid in the two-nucleon c.m. frame, $\overline{Q}$ is
evaluated for such a Fermi ellipsoid:
\beqa
\overline{Q} = \left\{ \begin{array}{c c c}
0                                                             &     & |\veck|<k_{min} \\
\frac{\gamma E^*(k)-E^*_{F}}{\gamma u | \veck | }     & for & k_{min}< | \veck | <k_{max} \\
1                                                             &     & k_{max} < |\veck|
\end{array} \right.
\label{subsec:SNM;eq:Pauli}
\eeqa
with $k_{min}=\sqrt{\kf^2- u^2 E_F^2}$,
$k_{max}=\gamma (u E_{F}+ \kf)$, and $u=|\vecu|$. The partially decoupled set of the one-dimensional
integral equations are solved
by the matrix inversions techniques of Haftel and Tabakin~\cite{haftel70}.
\\ \indent Due the anti-symmetry of these two-nucleon states the total isospin of the two-nucleon
system $({\rm I}=0,1)$ can be restored by the standard
selection rule
\beq
(-1)^{\rm L+S+I}=-1.
\label{selection}
\eeq
 The five independent partial wave
amplitudes in the helicity representation are obtained from the five independent on-shell amplitudes in the
$|JMLS>$-representation~\cite{erkelenz74}. After the summation over the total angular momentum one has
the five on-shell plane-wave helicity matrix elements
\beqa
<{\vecp} \lambda_1^{'} \lambda_2^{'}| T^{\rm I}(x)|
{\vecq} \lambda_1 \lambda_2>
= \sum\limits_{\rm J} \left( \frac{2{\rm J}+1}{4\pi}\right)
d^J_{\lambda^{'} \lambda}(\theta)
<\lambda_1^{'} \lambda_2^{'}| T^{\rm J,I}(\vecp,\vecq,x)|
\lambda_1 \lambda_2> ,
\nonumber \\
\label{tmatel1}
\eeqa
where $\theta$ is the scattering angle between $\bf q$ and $\bf p$ with
$\pabs=\qabs$. Furthermore, one has $\lambda=\lambda_1-\lambda_2$ and
$\lambda^{'}=\lambda_1^{'}-\lambda_2^{'}$.
The reduced rotation matrices $d^{\rm J}_{\lambda^{'} \lambda}(\theta)$
are those defined by  Rose \cite{rose57}.
\\ \indent Since we determine the $T$-matrix
elements in the two-particle c.m. frame, a representation with
covariant operators and Lorentz invariant amplitudes
in Dirac space is the most convenient way to Lorentz-transform the $T$-matrix
from the two-particle c.m. frame into  the nuclear matter rest frame
\cite{horowitz87}.  Some freedom in the choice of this
representation exists, because pseudoscalar ($ps$) and
pseudovector ($pv$) components
can not uniquely be disentangled for on-shell scattering. This ambiguity is minimized
by separating the leading order, i.e. the single-meson exchange, from
the full $T$-matrix. Therefore,  the contributions
stemming from the single-$\pi$ and-$\eta$ exchange are given in the complete
$pv$ representation. For the remaining part of the $T$-matrix, the $ps$ representation is chosen.
\\ \indent Taking the single nucleon 
momentum $\veck=(0,0,k)$ along the $z$-axis, then we have for the $nn$ and $pp$ channel contributions for the self-energy components in the $ps$ representation scheme
\beqa
\Sigma_s^{ij} (\veck) & = & 
\frac{1}{4}
\int_0^{\kfj} \frac{d^3\vecq}{(2 \pi)^3}  \frac{m^*_j}{E^*_{q,j}} 
[4 F^{ij}_{\rm S} - F^{ij}_{\rm \tilde{S}} - 4 F^{ij}_{\rm \tilde{V}} - 12
F^{ij}_{\rm \tilde{T}} + 4 F^{ij}_{\rm \tilde{A}} - F^{ij}_{\rm \tilde{P}}],
\label{subsec:PS;eq:s}
\eeqa
\beqa
\Sigma_o^{ij} (\veck) & = & 
 \frac{1}{4} 
\int_0^{\kfj} \frac{d^3\vecq}{(2 \pi)^3}  
[- 4 F^{ij}_{\rm V} + F^{ij}_{\rm \tilde{S}} - 2 F^{ij}_{\rm
    \tilde{V}}  - 2 F^{ij}_{\rm \tilde{A}} - F^{ij}_{\rm \tilde{P}}] ,
\label{subsec:PS;eq:o}
\eeqa
and
\beqa
\Sigma_v^{ij} (\veck) & = & 
\frac{1}{4} 
\int_0^{\kfj} \frac{d^3\vecq}{(2 \pi)^3} \frac{\vecq \cdot \veck}{|\veck|^2 E^*_{q,j}} 
[- 4 F^{ij}_{\rm V} + F^{ij}_{\rm \tilde{S}} - 2 F^{ij}_{\rm
    \tilde{V}}  - 2 F^{ij}_{\rm \tilde{A}} - F^{ij}_{\rm \tilde{P}}],
\label{subsec:PS;eq:v}
\eeqa
where $i=j=n$ or $i=j=p$, respectively.
In the complete $pv$ representation the $nn$ and $pp$ channel contributions to the self-energy components are given by 
\beqa
\Sigma_s^{ij} (\veck) & = & 
\frac{1}{4}
\int_0^{\kfj} \frac{d^3\vecq}{(2 \pi)^3}  \frac{m^*_j}{E^*_{q,j}} 
[4 g^{ij}_{\rm S} - g^{ij}_{\rm \tilde{S}} + 4 g^{ij}_{\rm A} 
+ \frac{m_j^{*2} + m_i^{*2} - 2 k^{* \mu} q^*_{\mu}}{(m_i^*+m_j^*)^2} g^{ij}_{\rm \widetilde{PV}}],
\eeqa
\beqa
\Sigma_o^{ij} (\veck) & = & 
+ \frac{1}{4} \int_0^{\kfj} \frac{d^3\vecq}{(2 \pi)^3}  
[ g^{ij}_{\rm \tilde{S}} - 2 g^{ij}_{\rm A} -  \frac{2 E^*_{k,i} (m_j^{*2}-k^{* \mu}
q^*_{\mu}) - E^*_{q,j} (m^{*2}_j - m^{*2}_i)}{E^*_{q,j} (m^*_i +
    m^*_j)^2}   g^{ij}_{\rm \widetilde{PV}}],
\eeqa
and
\beqa
\Sigma_v^{ij} (\veck) & = & 
\frac{1}{4} 
\int_0^{\kfj} \frac{d^3\vecq}{(2 \pi)^3} \frac{\vecq \cdot \veck}{|\veck|^2 E^*_{q,j}} 
[g^{ij}_{\rm \tilde{S}} - 2 g^{ij}_{\rm A} \nn
\\ & & - \frac{2 k^*_z (m^{*2}_j - k^{* \mu}
  q^*_{\mu} ) - q_z (m^{*2}_j - m^{*2}_i)}{q_z (m^*_i+m^*_j)^2} g^{ij}_{\rm \widetilde{PV}}],
\eeqa
where $i=j=n$ or $i=j=p$, respectively.
\\ \indent In short, the complete $pv$ representation is applied for $V_{\pi,\eta}$ and  the $ps$
representation is used for the $T_{Sub}=T - V_{\pi,\eta}$ to get the most favorable representation scheme, the subtracted $T$-matrix representation
scheme.
%%%%%%%%%%%%%%%%%%%%%%%%%%%%%%%%%%%%%%%%%%%%%%%%%%%%%%%%%%%%%%%%%%%%%%%%%%
\subsection{$np$ channel}
%%%%%%%%%%%%%%%%%%%%%%%%%%%%%%%%%%%%%%%%%%%%%%%%%%%%%%%%%%%%%%%%%%%%%%%%%%
\indent Since in isospin asymmetric nuclear matter one has to deal with two distinct nucleons states in the $np$ channel, this channel is more complicated than the $nn$ and $pp$ channel. Working with two distinct nucleons has consequences for the Thompson equation, the Pauli blocking operator, and the number of independent helicity matrix elements.

First, the Bonn potential~\cite{machleidt89} has to be made suitable to treat distinct particles in the medium. An important difference is that the neutrons and protons have unequal effective masses.
These distinct effective masses have to be accounted for, 
in particular in the evaluation of the potential matrix elements. The resulting one boson exchange (OBE) matrix elements can be found in appendix~\ref{app:pe}.

Second, the two-particle propagator $iG_iG_j$ in the BS equation has to be replaced by the Thompson propagator for the $np$ channel. The effective Thompson propagator for this channel is given by 
\beqa
g_{np}= i G_{n} G_{p} = \frac{m^*_n}{E^*_n} \frac{m^*_p}{E^*_p}
\frac{1}{\sqrt{s^*}-E^*_n-E^*_p + i \epsilon},
\eeqa
where $\sqrt{s^*}$ is the invariant mass.
\\ \indent In contrast to the five independent helicity matrix elements in the on-shell case for identical particles,
in the $np$ channel six helicity matrix elements are independent~\cite{tjon85a}. After the partial wave projection onto the $|JMLS>$-states, using an average direct-exchange contribution in the potential this time the Thompson equation  partially decouples  into two
subsystems of one-dimensional integral
equations: the coupled spin singlet-triplet states and the coupled triplet states (appendix~\ref{app:pwd}). To achieve the reduction to the
one-dimensional integral equations the Pauli operator
$Q$ has to be replaced by an angle-averaged Pauli operator
$\overline{Q}$ ~\cite{horowitz87}. 
However, the Pauli operator $Q$ for the $np$ channel has to be modified compared to the one in the $nn$ and $pp$ channel, since it has to be evaluated for
Fermi ellipsoids with different sizes. The result for the angle-averaged Pauli operator for the $np$ channel $\overline{Q}_{np}$  with a neutron excess is  
\beqa
\overline{Q}_{np}=\left\{ \begin{array}{c c c}
\Theta(\gamma u E_{Fn}-\gamma k_{Fn}) & & |\veck|<k_{min} \\
1/2 [\cos(\theta_p)-\cos(\theta_n)] \Theta(\theta_n-\theta_p) & for &
k_{min}<|\veck|<k_{max} \\
1 & & k_{max} < |\veck| \end{array} \right.
\eeqa
with $k_{min}=\gamma | u E_{Fn}- k_{Fn}|$,
$k_{max}=\gamma (u E_{Fn}+ k_{Fn})$, 
\beqa
\theta_p=\left\{ \begin{array}{c c c}
\arccos\left(\frac{\gamma E^*_p(k)-E^*_{Fp}}{\gamma |\veck| |\vecu|}\right) & for &
|\frac{\gamma E^*_p(k) -E^*_{Fp}}{\gamma |\veck| |\vecu|}|  \leq 1 \\
0 & & otherwise \end{array} \right., 
\eeqa
and
\beqa
\theta_n=\left\{ \begin{array}{c c c}
\arccos\left(\frac{E^*_{Fn} - \gamma E^*_n(k)}{\gamma |\veck| |\vecu|}\right) & for &
|\frac{E^*_{Fn} - \gamma E^*_n(k) }{\gamma |\veck| |\vecu|}|  \leq 1 \\
\pi & &  otherwise \end{array} \right. \quad .
\eeqa
\\ \indent 
Due to the additional independent helicity matrix element, we will have a sixth independent covariant in the $T$-matrix representation~\cite{tjon85a}. However, the problem is that we need to have a decomposition that reduces to the one used in the symmetric case. The general Lorentz representation of the nine invariants given in~\cite{tjon85a} fulfill this requirement. Leaving out the three redundant invariants in our case, the additionally constructed covariant
is defined as
\beqa
T^{{\rm I},dir}_6(\pabs,\theta,x)&=& {1\over 2} F_{\rm 6}^{\rm I}(\pabs,\theta,x)({\rm (\gamma_{\mu})_2 \cdot Q^{\mu}_1 - (\gamma_{\mu})_1 \cdot Q^{\mu}_2}) \nonumber \\ &=&{1\over 2} F_{\rm 6}^{\rm I}(\pabs,\theta,x) ({\rm \gamma_2 \cdot k + \gamma_1 \cdot k}), 
\eeqa
with  $Q^{\mu}_i=(p_i+q'_i)^{\mu}/2m$ for $i=1,2$ and $\veck=\vecp_1+\vecq'_1=-(\vecp_2+\vecq'_2)$ in the cm frame. The same sixth covariant is used in ref.~\cite{dejong98}, while the exchange sixth amplitude given in ref.~\cite{dejong98} does not contribute in the self energy components.
Therefore, one gets an additional term in the $np$ channel contribution to the neutron self energy components
\beqa
\Sigma^{np}_{s,6} (\veck) & = &
\frac{1}{4}
\int_0^{\kfp} \frac{d^3\vecq}{(2 \pi)^3}  \frac{m^*_p}{E^*_{q,p}} 
[4  \frac{k^{* \mu} q^*_{\mu} - m_p^{*2}  }{m_p^*}   F^{np}_{\rm 6} ],
\label{subsec:np6;eq:s}
\eeqa
\beqa
\Sigma^{np}_{o,6} (\veck) & = & 
\frac{1}{4} 
\int_0^{\kfp} \frac{d^3\vecq}{(2 \pi)^3}  
[4 m^*_p \frac{E^*_{k,n} - E^*_{q,p}}{E^*_{q,p}}  F^{np}_{\rm 6} ] ,
\label{subsec:np6;eq:o}
\eeqa
and 
\beqa
\Sigma^{np}_{v,6} (\veck) & = & 
\frac{1}{4} 
\int_0^{\kfp} \frac{d^3\vecq}{(2 \pi)^3} \frac{\vecq \cdot \veck}{|\veck|^2 E^*_{q,p}} 
[- 4 m^*_p \frac{k-q_z}{q_z} F^{np}_{\rm 6} ]
\label{subsec:np6;eq:v}
\eeqa
compared to the $nn$ and $pp$ channel. For the proton a similar additional term arises, where neutrons and protons are interchanged in eqs.~(\ref{subsec:np6;eq:s})-(\ref{subsec:np6;eq:v}).
In symmetric nuclear matter with equal effective masses for neutrons and protons, the coefficient of this  sixth independent amplitude vanishes, i.e. the familiar representation scheme with the five linearly independent covariants 
is obtained, as expected. 
\\ \indent Finally, the total neutron and proton self energies including all channels can be written as
\beqa
\Sigma^n = \Sigma^{nn} + \Sigma^{np} \quad; \Sigma^p =\Sigma^{pp} + \Sigma^{pn},
\eeqa
respectively.  
\section{Results and Discussion}
\label{sec:rd}
%%%%%%%%%%%%%%%%%%%%%%%%%%%%%%%%%%%%%%%%%%%%%%%%%%%%%%%%%%%%%%%%%%%%%%%%
\begin{figure}[!h]
\begin{center}
\includegraphics[width=0.9\textwidth] {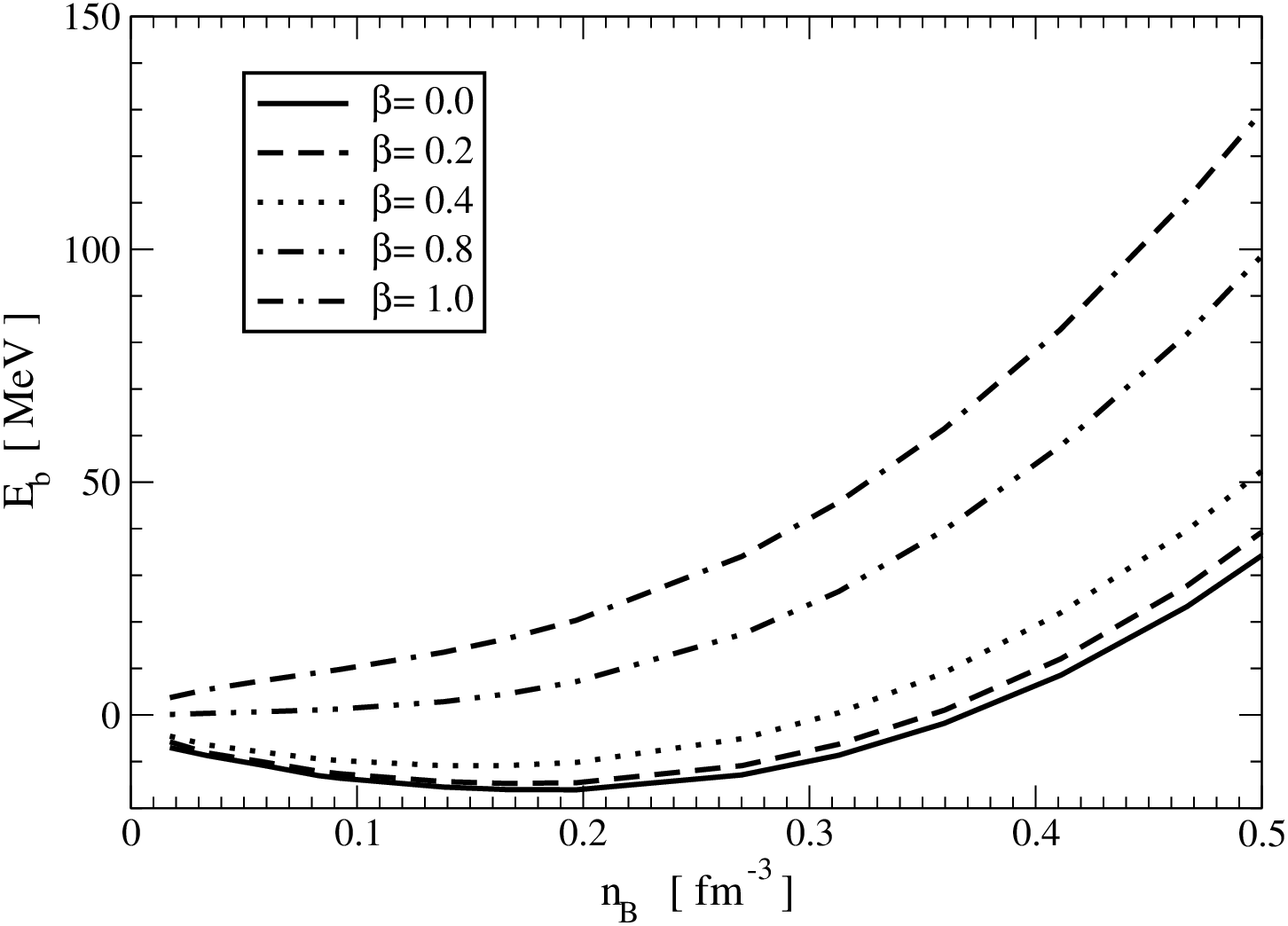}
\caption{Binding energy as a function of the baryon density.
\label{fig:Be}}
\end{center}
\end{figure}
%%%%%%%%%%%%%%%%%%%%%%%%%%%%%%%%%%%%%%%%%%%%%%%%%%%%%%%%%%%%%%%%%%%%%%%%%
In fig.~\ref{fig:Be} we present the results for the equation of
state for various values of the asymmetry parameter $\beta=(n_n-n_p)/n_B$ in the framework of the DBHF approach with  a sixth independent amplitude in the $np$ channel using the Bonn A
potential. The applied representation is the optimal representation so far, the subtracted
$T$-matrix representation. The two extreme cases are symmetric nuclear matter ($\beta=0.0$) and
neutron matter ($\beta=1.0$). The symmetric nuclear matter results and neutron matter results agree with those of refs.~\cite{gross99,vandalen04b}.
The binding energy curves for intermediate
values of $\beta$ lie between these two extreme curves and are slightly higher than in ref.~\cite{vandalen04b}.
In addition to that, the binding energy 
\beqa
E(n_B,\beta)=E(n_B)+E_{sym}(n_B) \beta^2 + {\cal O}(\beta^4)
\eeqa
shows a nearly quadratic dependence on the asymmetry parameter
$\beta$ as expected.
%%%%%%%%%%%%%%%%%%%%%%%%%%%%%%%%%%%%%%%%%%%%%%%%%%%%%%%%%%%%%%%%%%%%%%%%
\begin{figure}[!h]
\begin{center}
\includegraphics[width=0.9\textwidth] {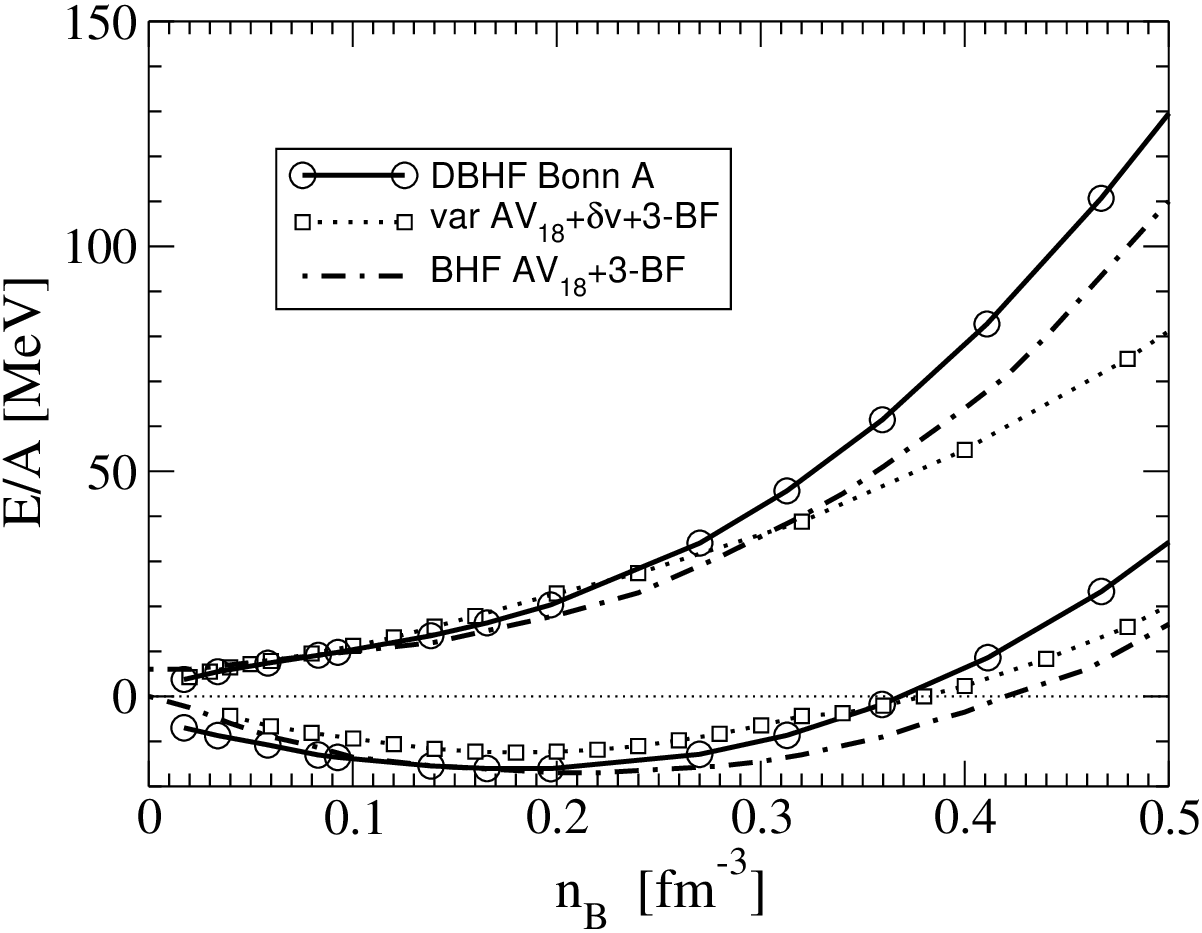}
\caption{Comparison of several EoSs from ab initio calculations, i.e. 
the present approach (solid), a nonrelativistic BHF~\cite{lejeune00} (dashed-dotted) and a variational calculation~\cite{akmal98} (dotted).
\label{fig:BeLIT}}
\end{center}
\end{figure}
%%%%%%%%%%%%%%%%%%%%%%%%%%%%%%%%%%%%%%%%%%%%%%%%%%%%%%%%%%%%%%%%%%%%%%%%%

Fig.~\ref{fig:BeLIT} compares our prediction for the binding energy to the ones of other 
microscopic many-body approaches, the variational calculations from~\cite{akmal98} and the nonrelativistic BHF calculation from~\cite{lejeune00}, at symmetric nuclear matter (below zero) and pure neutron matter 
(above zero). The variational calculation is based on the high precision phenomenological Argonne $V_{18}$~\cite{wiringa95} two-nucleon interaction and includes UIX three-body forces~\cite{pudliner95} as well as relativistic boost correction denoted by $\delta v$~\cite{akmal98}. Also the nonrelativistic BHF calculation~\cite{lejeune00} is based on the phenomenological Argonne $V_{18}$~\cite{wiringa95}. Furthermore, it includes a microscopic three-body force deduced from the meson-exchange current approach~\cite{lejeune00}.

The first observation that becomes evident from fig.~\ref{fig:BeLIT} is that in nuclear matter both, the BHF and DBHF, calculations lead to more binding than the variational calculation. 
However, in all three cases the EoS of nuclear matter can be characterized as "soft\'', at least at moderate densities up to about three times saturation density. The prediction of a soft EoS is the general outcome of a microscopic  many-body calculation.
Recent Quantum Monte Carlo calculations for symmetric nuclear matter~\cite{qmc} show the same tendency. It should be noticed that this observation is supported by corresponding observables extracted from heavy ion reactions, where supranormal densities up to about three times saturation density are probed. Heavy ion data for tranverse flow~\cite{stoicea04} or from kaon production~\cite{sturm01} support the picture of a soft EoS in symmetric nuclear matter.

In neutron matter the variational calculations are less stiff, in particular at high-density neutron matter, than our DBHF calculations, whereas the nonrelativistic BHF calculation lies in between these two approaches. However, up to $1.5$ times saturation density for neutron matter and symmetric nuclear matter the three approaches show a quite reasonable agreement. This fact indicates that these are the density ranges which are at present reasonable well controlled by state-of-the-art many-body calculations.

The high density behavior of the EoS, in particular that of the neutron matter EoS, can be constructed by astrophysical observables~\cite{vandalen06b}. The recent observation of the at present heaviest compact star, a binary pulsar of $2.1 \pm 0.2 M_\odot$ (1$\sigma$ level)~\cite{nice05} rules out very soft neutron matter EoSs. However, all three EoSs shown in fig.~\ref{fig:BeLIT} fulfill this constraint since they yield maximum neutron star masses between $2.2 \div 2.3 M_\odot$.

%%%%%%%%%%%%%%%%%%%%%%%%%%%%%%%%%%%%%%%%%%%%%%%%%%%%%%%%%%%%%%%%%%%%%%%%
\begin{figure}[!h]
\begin{center}
\includegraphics[width=0.9\textwidth] {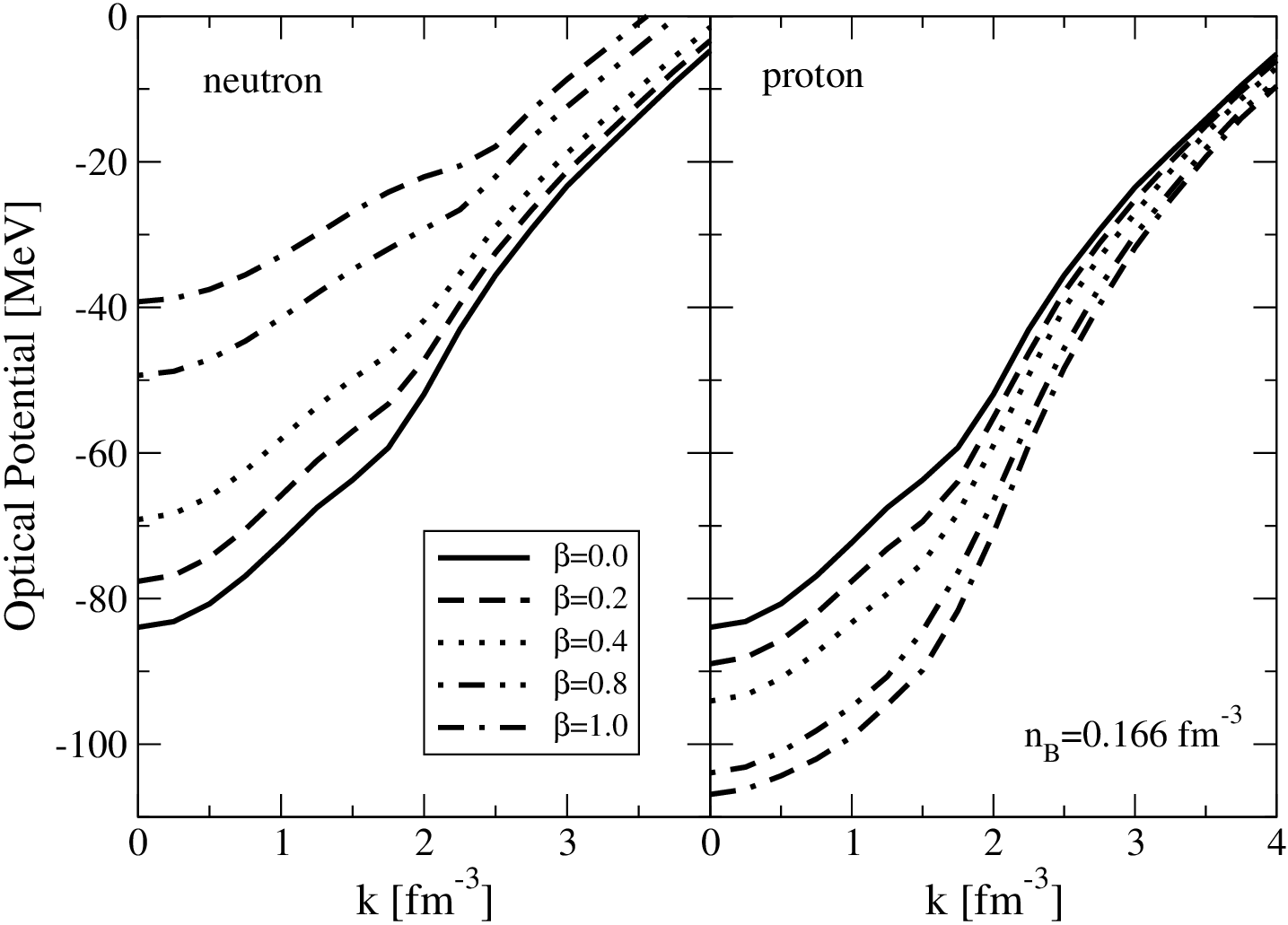}
\caption{The neutron and proton optical potential in neutron rich matter
as a function of the
momentum $k=|\veck|$.
\label{fig:aUopt}}
\end{center}
\end{figure}
%%%%%%%%%%%%%%%%%%%%%%%%%%%%%%%%%%%%%%%%%%%%%%%%%%%%%%%%%%%%%%%%%%%%%%%%%

In fig.~\ref{fig:aUopt}
the neutron and proton optical potentials are plotted
as a function of the
momentum $k=|\veck|$ for various values
of the asymmetry parameter $\beta = (n_n - n_p)/n_B$ at a fixed nuclear density of 
$n_B = 0.166 \ \textrm{fm}^{-3}$. The depth of neutron optical potential
decreases with increasing asymmetry, whereas the depth of the proton optical potential
shows the opposite behavior. Furthermore, the steepness of the neutron optical potential decreases
with increasing asymmetry parameter $\beta$, whereas the opposite behavior is found
in the proton case. Compared to ref.~\cite{vandalen05b} the neutron optical potential remains almost unaltered. In contrast,
the proton optical potential lies a bit  lower and is somewhat steeper as compared to ref.~\cite{vandalen05b}.

The isovector optical  potential $U_{iso}= \frac{U_n - U_p}{2 \beta}$ strongly depends on density and momentum. This
optical potential in neutron-rich matter initially stays constant 
and then decreases strongly with increasing momentum. Furthermore, the isovector optical potential is almost independent
of the asymmetry parameter $\beta$. This behavior can also be observed in refs.~\cite{vandalen04b,vandalen05a,vandalen05b}. 
Since the proton optical potential lies a bit lower, the isovector optical
potential  at
$k=0$ is slightly higher than in refs.~\cite{vandalen04b,vandalen05a,vandalen05b}.
However, the optical isovector
potential at nuclear density $n_B = 0.166 \ \textrm{fm}^{-3}$ at
$k=0$ is still in good agreement with the empirical value of 22 - 34
MeV~\cite{li04}.

%%%%%%%%%%%%%%%%%%%%%%%%%%%%%%%%%%%%%%%%%%%%%%%%%%%%%%%%%%%%%%%%%%%%%%%%
\begin{figure}[!h]
\begin{center}
\includegraphics[width=0.9\textwidth] {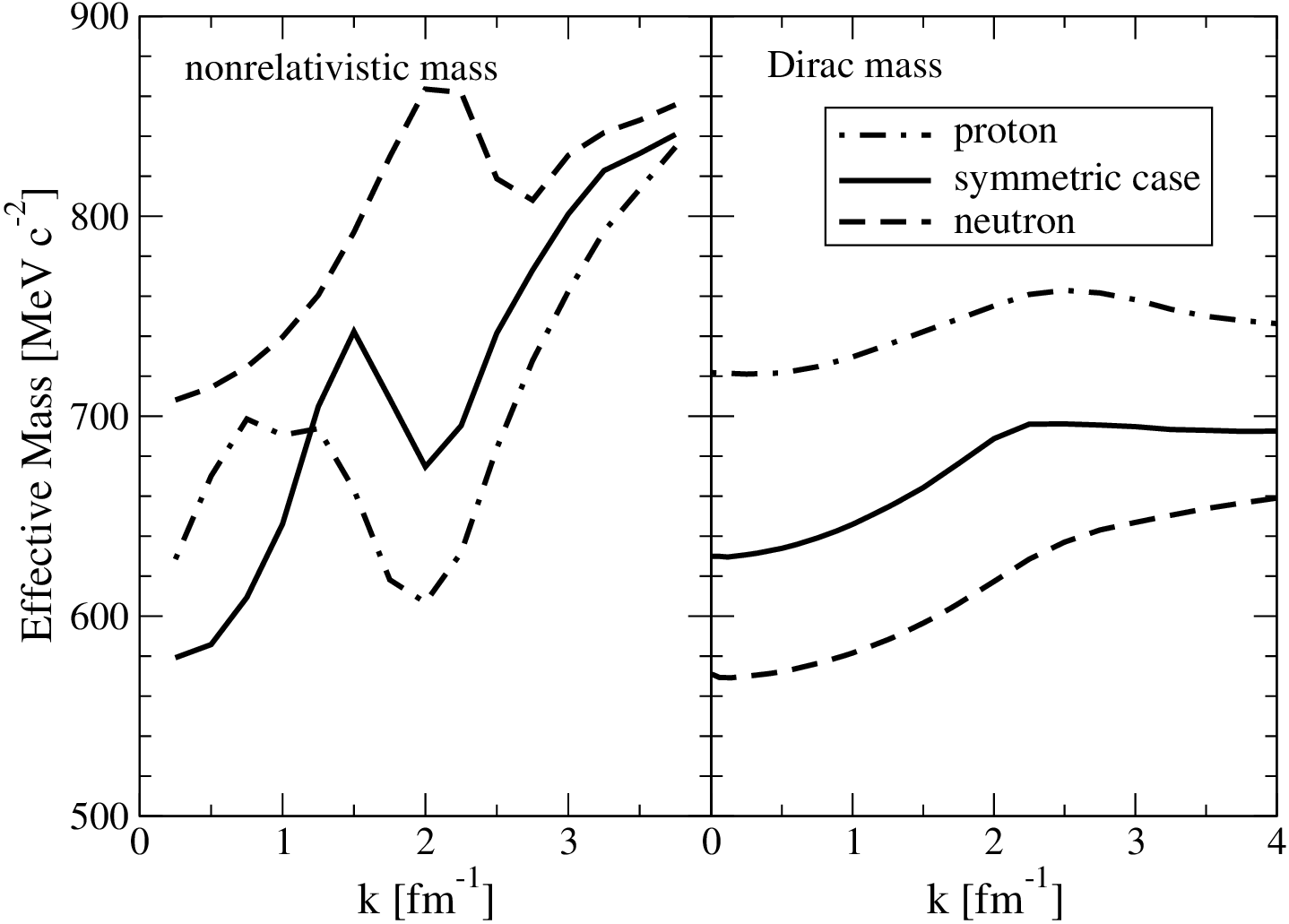}
\caption{Neutron and proton effective mass as a function of the
momentum $k=|\veck|$ in neutron matter at fixed nuclear density
$n_B = 0.166 \quad \textrm{fm}^{-3}$. In addition,
the effective mass in symmetric nuclear matter is given.
\label{fig:asplit}}
\end{center}
\end{figure}
%%%%%%%%%%%%%%%%%%%%%%%%%%%%%%%%%%%%%%%%%%%%%%%%%%%%%%%%%%%%%%%%%%%%%%%%%
An interesting issue is the proton-neutron mass splitting in
neutron-rich matter, which has in detail been discussed in refs.~\cite{vandalen05a,vandalen05b}. One should keep in mind that different definitions of the effective mass exist, which are often compared and sometimes even mixed up: the nonrelativistic mass and the relativistic Dirac mass.
In fig.~\ref{fig:asplit} the nonrelativistic and Dirac
effective mass of the neutron and proton are compared for $\beta=1$, i.e.
neutron matter. Our DBHF calculations
based on projection techniques predict a mass splitting of
$m^*_{D,n} < m^*_{D,p}$ in neutron-rich
matter.
However, the nonrelativistic mass derived from our DBHF approach shows the
opposite behavior. This opposite behavior to the relativistic
Dirac mass, i.e. $m^*_{NR,n} > m^*_{NR,p}$, is in agreement with
the results from nonrelativistic BHF calculations
\cite{muether02}. This difference between the Dirac mass splitting and the nonrelativistic mass
splitting is not surprising, since these masses are based on completely different physical concepts.
The relativistic Dirac mass 
is defined through the scalar part of the nucleon self-energy in the 
Dirac field equation which is absorbed into the effective mass~(\ref{subsec:SM;eq:dirac}). 
On the other hand, the nonrelativistic mass parameterizes
the momentum dependence of the single particle potential. 

In this context we want to note that, in contrast to the non-relativisitc 
mass $m^*_{NR}$,  the momentum dependence of the Dirac mass $m^*_D$ is smooth 
and still moderate. This fact is important to justify the reference spectrum 
approximation, i.e. the usage of an momentum independent effecive Dirac mass 
${\tilde m}^*_F$ for the evaluation of the in-medium spinor basis (\ref{spinor}), 
the Thompson propagator and the potential matrix elements (see Appendix).

%%%%%%%%%%%%%%%%%%%%%%%%%%%%%%%%%%%%%%%%%%%%%%%%%%%%%%%%%%%%%%%%%%%%%%%%
\begin{figure}[!t]
\begin{center}
\includegraphics[width=0.9\textwidth] {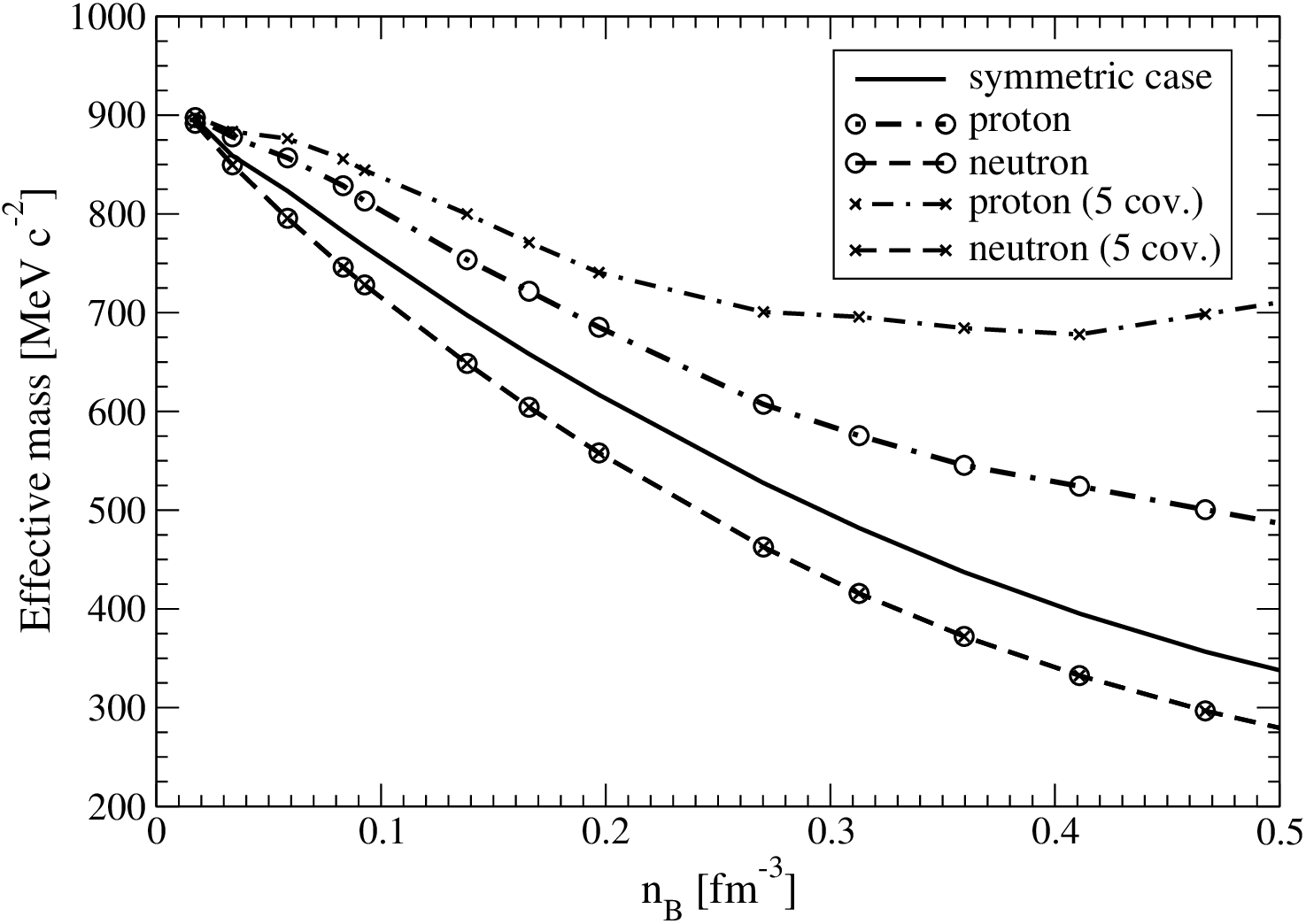}
\caption{Neutron and proton effective mass as a function of the
density in neutron matter. In addition, the effective mass in symmetric nuclear matter is given. These results are compared to DBHF calculations with average mass assumption in the $np$ channel (5 cov.) of refs.~\cite{vandalen04b,vandalen05a,vandalen05b}.
\label{fig:cov5}}
\end{center}
\end{figure}
%%%%%%%%%%%%%%%%%%%%%%%%%%%%%%%%%%%%%%%%%%%%%%%%%%%%%%%%%%%%%%%%%%%%%%%%
In fig.~\ref{fig:cov5} the neutron and proton effective Dirac masses are plotted as a
function of the baryon density $n_B$ for pure neutron matter. Of course, a
strong density dependence can be observed. In addition, one has a Dirac mass splitting of
$m^*_{D,n} < m^*_{D,p}$ in the whole density range. RMF field theories with the isovector $\rho$ and $\delta$ mesons included predict
the same Dirac mass splitting. When only the $\rho$-meson is included, the RMF theory predicts equal masses. Hence, 
the $\delta$ meson is responsible for the mass splitting in RMF theory.  

Furthermore, in fig.~\ref{fig:cov5} our results for the neutron and proton effective Dirac mass in pure neutron matter are compared to those from refs.~\cite{vandalen04b,vandalen05a,vandalen05b}, where only 5 covariants were used in the $np$ channel.
The neutron effective mass remains practically unaffected, whereas the proton mass experiences a sizable reduction. These results are easy to understand. The neutron self energy consists of
a $nn$ and a $np$ part. Hence, the $nn$ part becomes dominant for a vanishing proton fraction. The proton self energy consists of a $pp$ and a $np$ part. In the limit of a vanishing proton fraction, the $np$ interaction becomes dominant. Therefore, the proton properties, e.g. the proton effective mass, are especially sensitive for the treatment of the $np$ channel in neutron rich matter.
%%%%%%%%%%%%%%%%%%%%%%%%%%%%%%%%%%%%%%%%%%%%%%%%%%%%%%%%%%%%%%%%%%%%%%%%%
\section{Relation to relativistic mean field theory.}
\subsection{DBHF self energy components}
At present full Brueckner calculation are still too complex to allow an application to finite nuclei.
However, within the framework of density dependent mean field theory
effective density depend coupling functions can be  obtained from the Brueckner self-energy components. Such coupling functions parameterize the correlations of the 
$T$-matrix in a handable way and can be applied to finite nuclei 
within the framework of DDRMF 
theory~\cite{fuchs95}. In contrast to standard RMF models, the meson-baryon vertices are density dependent.
As a consequence, rearrangement contributions in the baryon field equations occur. These rearrangement contributions should be taken into account and are essential to satisfy energy-momentum conservation and thermodynamic consistency
in this density dependent mean field theory.  

In order to properly parameterize the isospin dependence of the self-energy components, the coupling functions must be based on four different channels: scalar isoscalar, vector isoscalar, scalar
isovector, and vector isovector channel. In RMF theory these channels correspond to phenomenological exchange bosons, i.e. the $\sigma$, $\omega$, $\delta$, and $\rho$ mesons. The
effective coupling constants are then given by  
\beqa
\left(\frac{g_{\sigma}(n_B,\beta)}{m_{\sigma}}\right)^2 = - \frac{1}{2}
\frac{\Sigsp(\kfp)  + \Sigsn(\kfn)}{n_s}, \label{eq:ss}\\  
\left(\frac{g_{\omega}(n_B,\beta)}{m_{\omega}}\right)^2 = - \frac{1}{2}
\frac{\Sigop(\kfp)  + \Sigon(\kfn)}{n_B}, \label{eq:vw} \\  
\left(\frac{g_{\delta}(n_B,\beta)}{m_{\delta}}\right)^2 = - \frac{1}{2}
\frac{\Sigsp(\kfp)  - \Sigsn(\kfn)}{n_{s3}}, \label{eq:isd}\\
\left(\frac{g_{\rho}(n_B,\beta)}{m_{\rho}}\right)^2 = - \frac{1}{2}
\frac{\Sigop(\kfp)  - \Sigon(\kfn)}{n_3},  \label{eq:ivr}    
\eeqa
with $n_s=n_{sp}+n_{sn}$, $n_B=n_p+n_n$, $n_{s3}=n_{sp}-n_{sn}$, and $n_3=n_p-n_n$,
where 
\beqa
n_{si}=\frac{2}{(2\pi)^3} \int_0^{k_{Fi}} d\veck \frac{m^*_i}{\sqrt{{m^*_i}^2+k^2}}
\eeqa 
and
\beqa
n_i=\frac{2}{(2\pi)^3} \int_0^{k_{Fi}} d\veck = \frac{k_{Fi}^3}{3 \pi^2} 
\eeqa
are, respectively, the scalar and vector density of particle $i(=n,p)$.
\begin{figure}[!h]
\begin{center}
\includegraphics[width=0.9\textwidth] {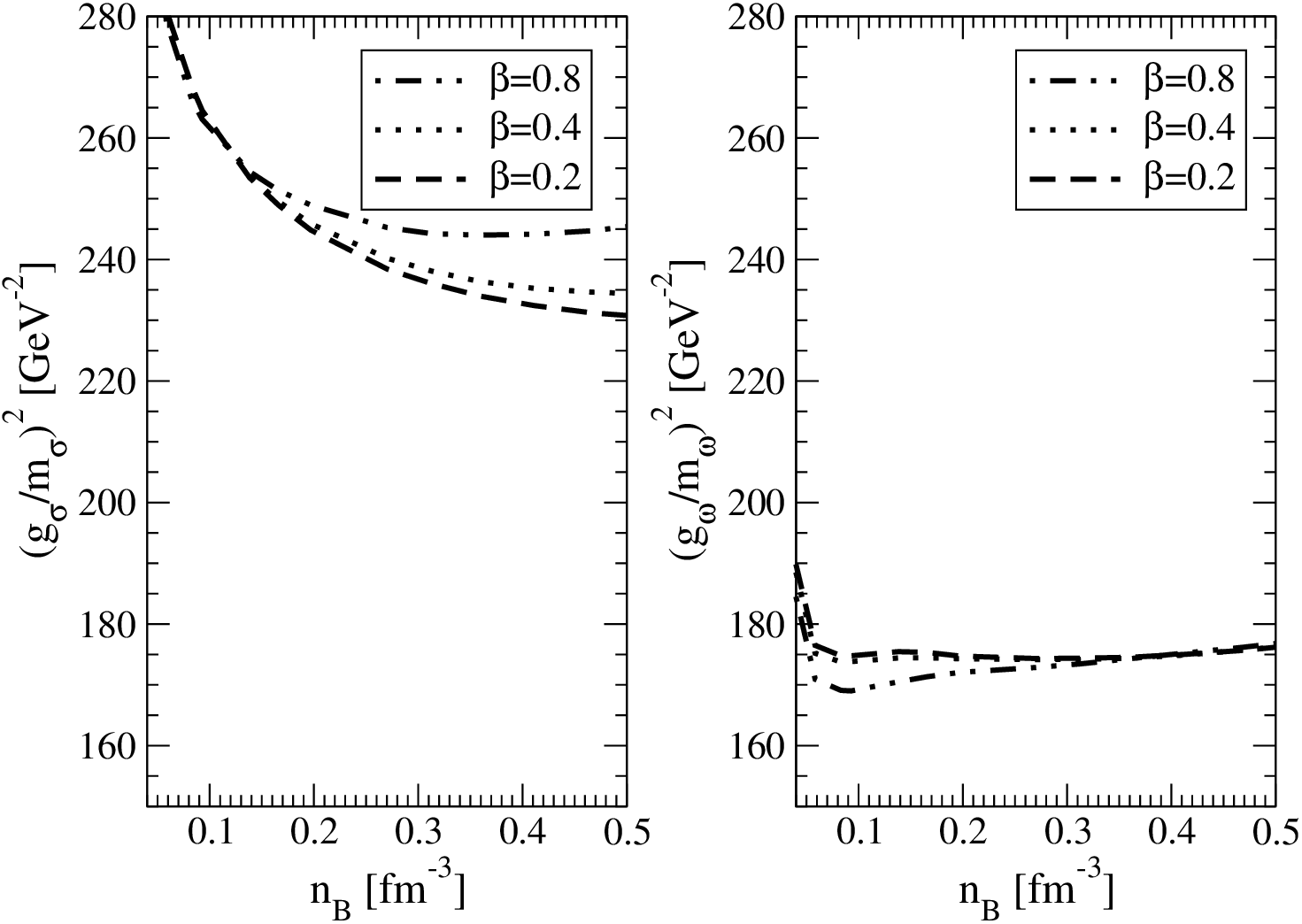}
\caption{The isoscalar scalar ($g_\sigma$) and vector ($g_\omega$) effective coupling functions are plotted as a function of the baryon density for different values of the  asymmetry parameter $\beta$. \label{fig:sigmaomega}}
\end{center}
\end{figure}
The results for the isoscalar and isovector coupling constants are
plotted in figs.~\ref{fig:sigmaomega}
\begin{figure}[!h]
\begin{center}
\includegraphics[width=0.9\textwidth] {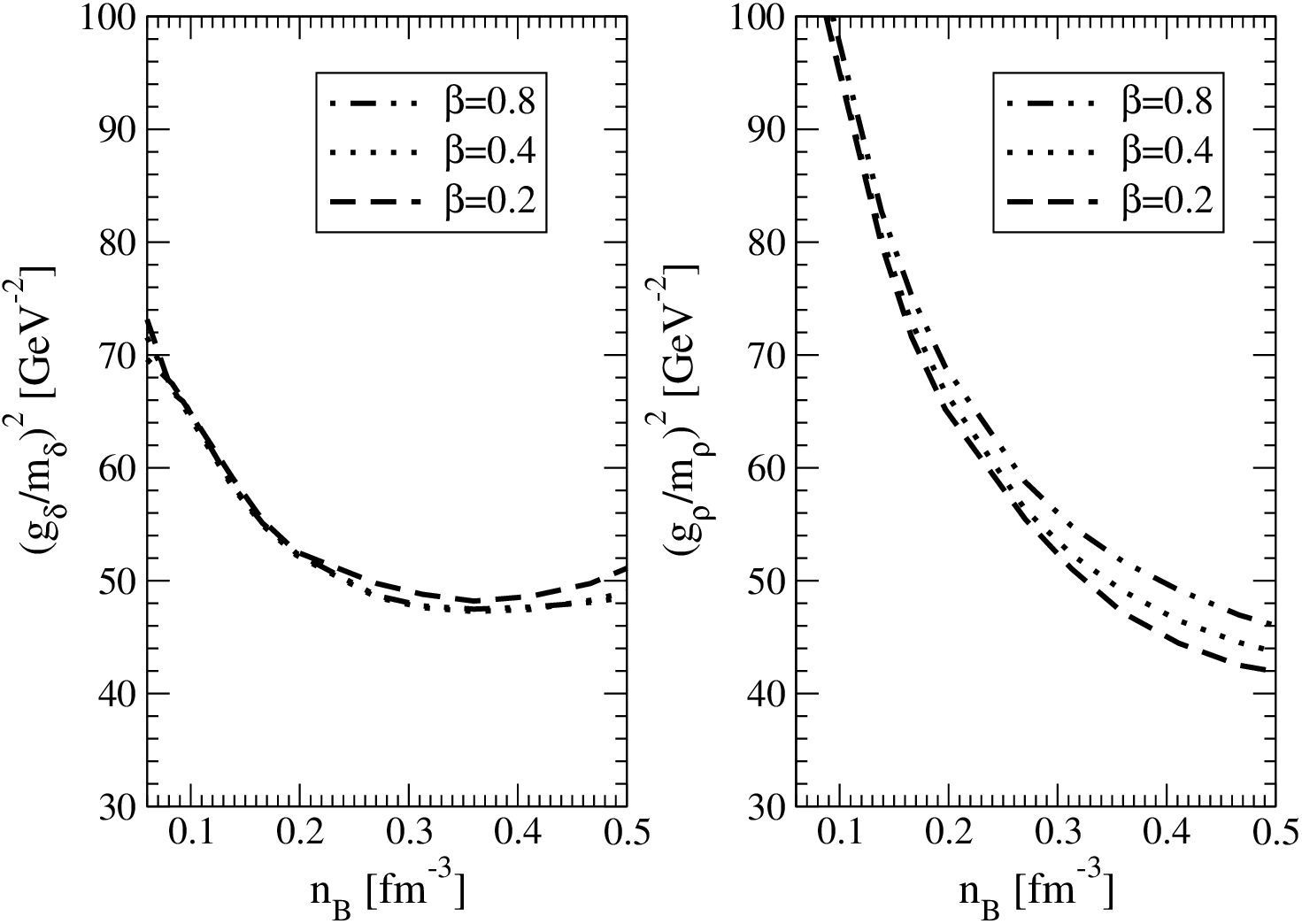}
\caption{The isovector scalar ($g_\delta$) and vector ($g_\rho$) effective coupling functions are plottted as a function of the baryon density for different values of the  asymmetry parameter $\beta$.  \label{fig:deltarho}}
\end{center}
\end{figure}
and~\ref{fig:deltarho},
respectively.  The strength of the isoscalar coupling functions
decreases as the density increases.  At low densities, both the scalar 
$g_\sigma$  and the vector isoscalar coupling $g_\omega$ show a strong decrease
with increasing density. However, at higher densities the vector coupling stays more or less
constant. The strength in the isovector channel is small
compared to that in the isoscalar channel. Furthermore, compared to ref.~\cite{vandalen04b} the dependence on the proton fraction for isovector strength is strongly reduced. 
\begin{figure}[!h]
\begin{center}
\includegraphics[width=0.9\textwidth] {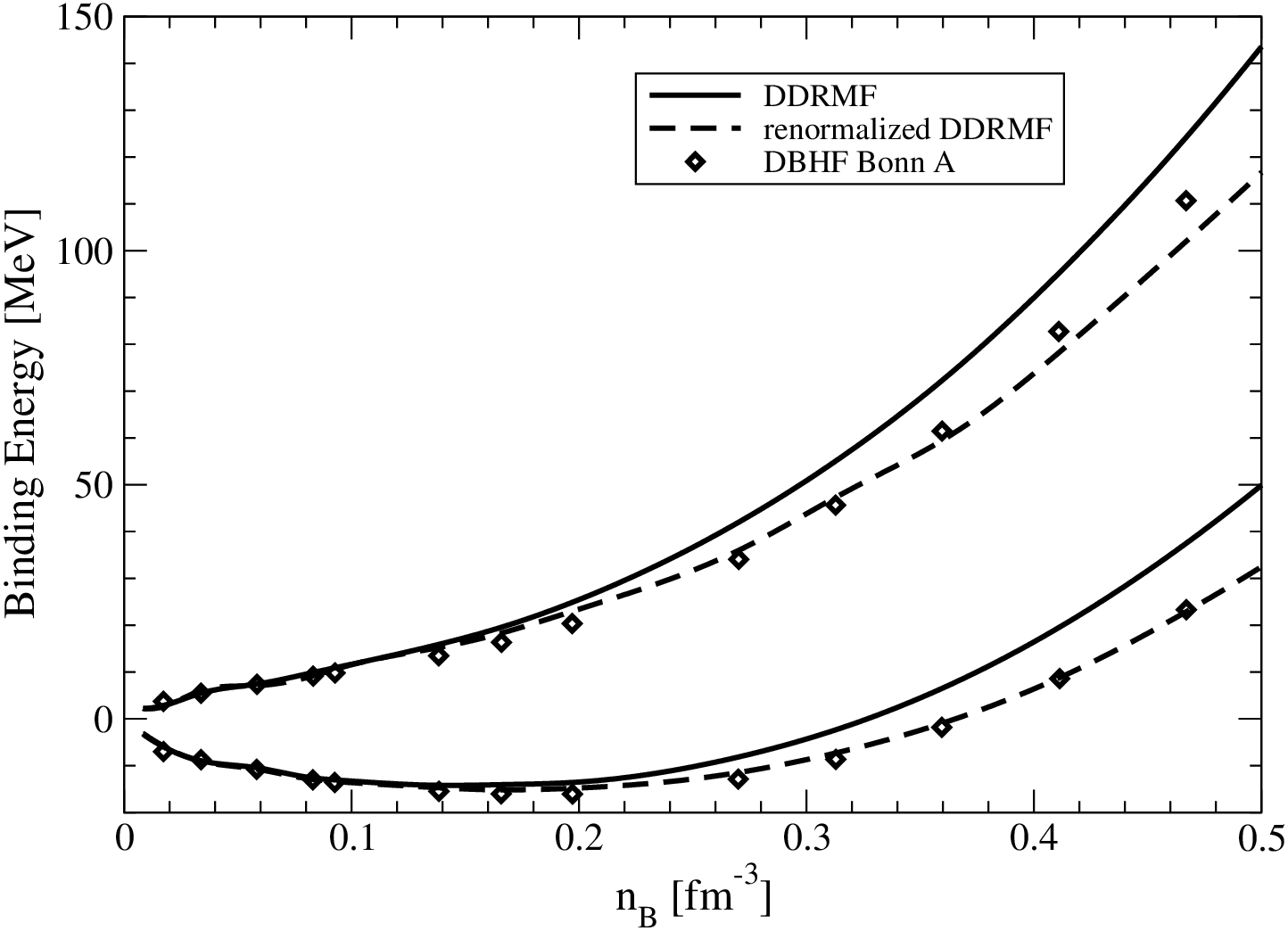}
\caption{The DBHF EoS is compared to DDRMF EoS and the renormalized DDRMF EoS.
\label{fig:EbindDBHF}}
\end{center}
\end{figure}

We can directly use these density dependent coupling functions in a RMF theory for infinite nuclear matter. 
For practical purposes and to keep the DDRMF functional as simple as possible we ignore in the following the weak isospin dependence and assume only a density dependence in the coupling functions of eqs.~(\ref{eq:ss})-(\ref{eq:ivr}). In fig.~\ref{fig:EbindDBHF} the binding energy determined from this RMF theory is compared to our DBHF results for neutron matter and symmetric nuclear matter.
The binding energy in RMF theory is given by 
\beqa
E/A & = & \frac{2}{(2 \pi)^3} \sum_{i=n,p} \int_{\Theta_{Fi}} d^3k E^*_i(k)+\frac{1}{2} 
\Bigg[\left(\frac{g_{\sigma}(n_B)}{m_{\sigma}}\right)^2 n_s^2+\left(\frac{g_{\omega}(n_B)}{m_{\omega}}\right)^2 n_B^2 \nonumber \\
& & + \left(\frac{g_{\delta}(n_B)}{m_{\delta}}\right)^2 n_{s3}^2+\left(\frac{g_{\rho}(n_B)}{m_{\rho}}\right)^2 n_3^2\Bigg], 
\label{eq:EADDRMF}
\eeqa
where the density dependent couplings 
\beqa
\left(\frac{g_{\alpha}(n_B)}{m_{\alpha}}\right)^2, \alpha  \in \{\sigma,\omega,\delta,\rho\}
\eeqa
are obtained from eqs.~(\ref{eq:ss})-(\ref{eq:ivr}) using the data for $\beta=0.2$. No rearrangement terms are present in eq. (\ref{eq:EADDRMF}), since rearrangement contributions do not contribute at the level of the binding energy~\cite{fuchs95}.
In RMF theory the integral for the kinetic energy can be evaluated and leads to the analytical expression
\beqa
\frac{2}{(2 \pi)^3} \sum_{i=n,p} \int_{\Theta_{Fi}} d^3k E^*_i(k)= \sum_{i=n,p}  \left[\frac{3}{4} E_{Fi} n_i+ \frac{1}{4} m^*_i n_{s,i}\right]
\eeqa
with the Fermi energy $E_{Fi}=\sqrt{k_{Fi}^2+m_i^{*2}}$. The effective mass contains the contributions of the two scalar mesons. Through the different coupling to the isovector $\delta$-meson it accounts for the proton-neutron mass splitting, i.e. 
\beqa
m^*_{n/p}=M-\left(\frac{g_{\sigma}(n_B)}{m_{\sigma}}\right)^2 n_s \pm \left(\frac{g_{\delta}(n_B)}{m_{\delta}}\right)^2 n_{s3}
\eeqa
Comparing the original DBHF EoS in Fig~(\ref{fig:EbindDBHF}) for the DDRMF EoS based on the parameterization (\ref{eq:ss})-(\ref{eq:ivr}), one observes clear deviations of the two approaches, both for symmetric as well as for neutron matter. This suggests that the density dependent coupling functions should be extracted more carefully as has been done in the "naive" definition (\ref{eq:ss})-(\ref{eq:ivr}). With other words, an accurate reproduction of the DBHF EoS requires a renormalization of the coupling functions which includes the contributions from Fock terms in a more consistent way.
%%%%%%%%%%%%%%%%%%%%%%%%%%%%%%%%%%%%%%%%%%%%%%%%%%%%%%%%%%%%%%%%%%%%%%%%%
\subsection{Renormalized self energy components}
%%%%%%%%%%%%%%%%%%%%%%%%%%%%%%%%%%%%%%%%%%%%%%%%%%%%%%%%%%%%%%%%%%%%%%%%
The fact that renormalization is required when DBHF results are mapped on RMF theory can easily be seen from the DBHF binding energy,
\beqa
E/A&=&\frac{2}{(2 \pi)^3} \sum_{i=n,p} \int_{\Theta_{Fi}} d^3k \left[E^*_i(k)-\Sigma_{o,i}
-\frac{1}{2} \Sigma_{s,i} \frac{m^*_i}{E^*_i}+\frac{1}{2} \frac{\Sigma_{\mu,i} k^{*\mu}}{E^*_i}\right]. 
\label{eq:EADBHF}
\eeqa
The two essential differences between DBHF and RMF concerning the structure of the self-energy, respectively the mean field,
are firstly, that the DBHF self-energies carry an explicit momentum dependence and, secondly, the appearance of a nonvanishing spatial contribution $\Sigma_V$, see eqs.~(\ref{subsec:SM;eq:self1})-(\ref{subsec:SM;eq:dirac2}). 
Both features should be taken into account as accurate as possible when DBHF results are parameterized in terms of RMF theory. 
The $\Sigma_V$ component originates from Fock exchange contributions which are not present in RMF theory. For an accurate reproduction of the DBHF energy functional the spatial $\Sigma_V$ component has to be included in proper way.
Firstly, the $\Sigma_V$ component can be absorbed into the effective mass according to eq.~(\ref{subsec:SM;eq:redquantity}) and this reduced effective mass has to be identified with RMF effective mass, i.e.
\beqa
\tilde{m}^{*}_i=\frac{M + \Sigsi(k_{Fi})}{1+\Sigvi(k_{Fi})}=M+\SigsDDRMFi. 
\label{eq:meffDBHF}
\eeqa
This leads to the renormalized scalar self energy component 
\beqa
\SigsDDRMFi=\frac{\Sigsi(k_{Fi})-M \Sigvi(k_{Fi})}{1+\Sigvi(k_{Fi})}. 
\eeqa
However, the DBHF energy functional of eq.~(\ref{eq:EADBHF}) has additional terms
compared to the DDRMF energy functional of eq.~(\ref{eq:EADDRMF}).
In the same way, however, then using the energy density instead of the effective mass, the following expression for the normalized vector self energy component is obtained
\beqa
\SigoDDRMFi=\Sigoi(k_{Fi})-\frac{\Sigvi(k_{Fi}) [3 E_{Fi} n_i +  \tilde{m}^*_i n_{s,i}]}{4 n_i}.
\eeqa
These renormalized self energy components are now inserted into eqs.~(\ref{eq:ss})-(\ref{eq:ivr}) to obtain the renormalized density dependent coupling functions.
By this procedure all terms which contribute to the DBHF energy functional are taken into account in the correspondingly constructed DDRMF functional. However, the explicit momentum dependence in eq.~(\ref{eq:EADBHF}) can not so easily be transferred to the RMF theory which leads still to slight deviation of the corresponding energy functionals.
A possibility would be to perform a Taylor expansion of the self-energy components in terms of the momentum~\cite{hofmann01}.
Since the intrinsic momentum dependence of the DBHF self-energy components is generally weak~\cite{gross99,vandalen04b} we neglect such additional correction terms.
%%%%%%%%%%%%%%%%%%%%%%%%%%%%%%%%%%%%%%%%%%%%%%%%%%%%%%%%%%%%%%%%%%%%%%%%%
%%%%%%%%%%%%%%%%%%%%%%%%%%%%%%%%%%%%%%%%%%%%%%%%%%%%%%%%%%%%%%%%%%%%%%%%
\begin{figure}[!h]
\begin{center}
\includegraphics[width=0.9\textwidth] {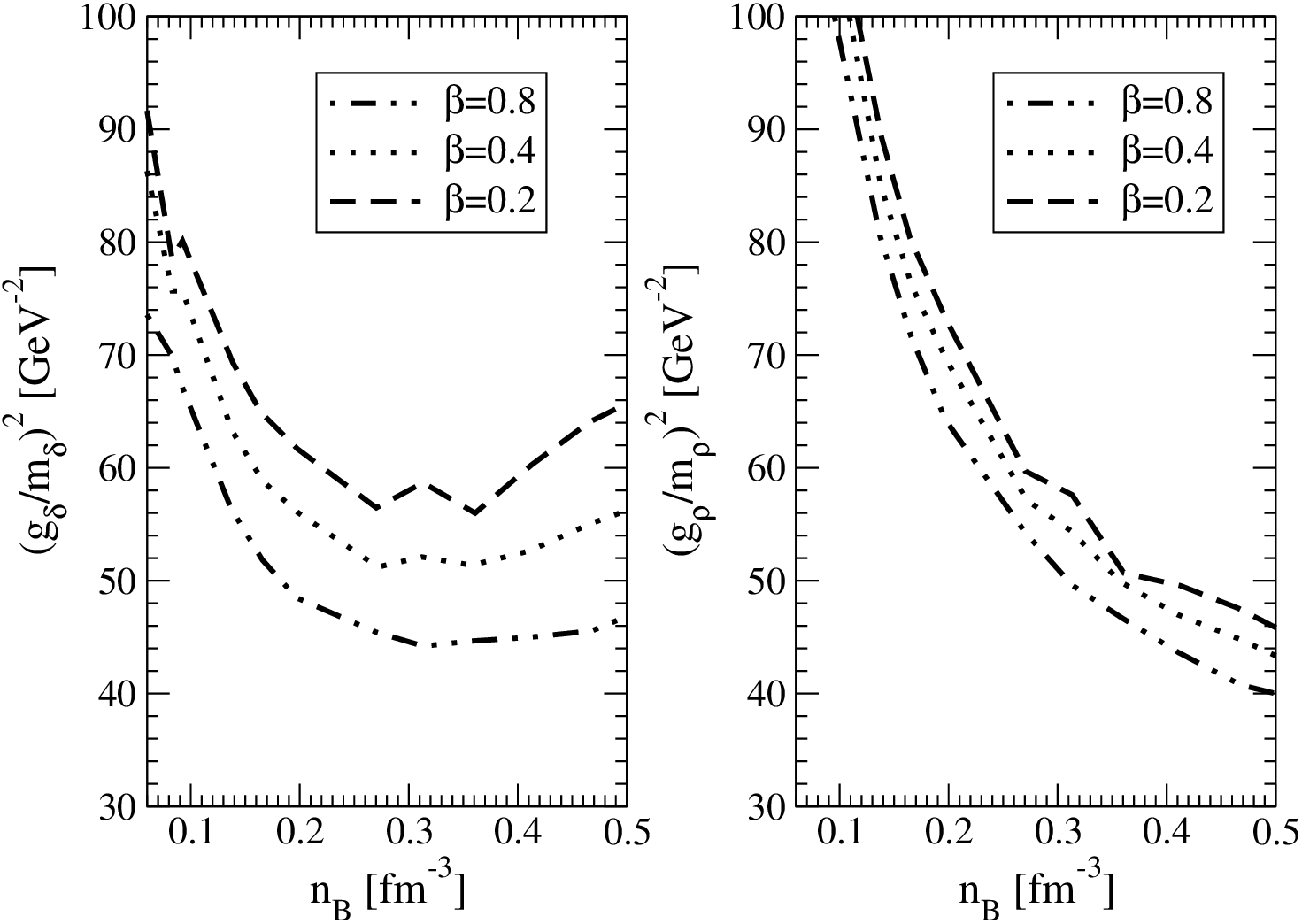}
\caption{The renormalized  isovector scalar ($g_\delta$) and vector ($g_\rho$) effective coupling functions are plotted 
as a function of the baryon density for different values of the asymmetry parameter
$\beta$.  
\label{fig:isovectorRMF}}
\end{center}
\end{figure}
%%%%%%%%%%%%%%%%%%%%%%%%%%%%%%%%%%%%%%%%%%%%%%%%%%%%%%%%%%%%%%%%%%%%%%%%%
The new renormalized isoscalar density dependent coupling functions are reduced by an amount of 15-20 MeV compared to the corresponding nonrenormalized coupling functions in fig.~\ref{fig:sigmaomega}, but the qualitative behavior is very similar. The small isospin dependence of the isoscalar strength is almost insensitive to the renormalization. 
The renormalized isovector  $\rho$ and $\delta$ meson coupling functions are shown in fig.~\ref{fig:isovectorRMF}.
The density dependence of the renormalized isovector coupling functions is similar to those of fig.~~\ref{fig:deltarho}. However, the $\beta$-dependence is now more pronounced.

These renormalized density coupling functions are applied in RMF theory. Again the weak isospin dependence of these coupling functions is ignored. Therefore, we use again the data for $\beta=0.2$ which corresponds approximately to the asymmetry in an $Au$ nucleus.
\begin{table}[!h]
\begin{tabular}{|c|c|c|c|}
\hline
model & $n_{sat} [\textrm{fm}^{-3}]$ & $k_F [\textrm{fm}^{-1}]$ & $E/A$ [MeV]  \\
\hline
DDRMF & 0.143 & 1.28 & -14.22 \\
renormalized DDRMF & 0.168 & 1.36 & -15.18\\
DBHF & 0.181 & 1.39 & -16.15 \\
\hline
\end{tabular}
\caption{Saturation properties of the DBHF model and the correspondingly constructed relativistic mean field functionals. \label{tab:sat}}
\end{table}
From table~\ref{tab:sat}, one can see that the saturation density is shifted to lower densities and the binding energy of the saturation point is weaker in renormalized DDRMF theory compared to the original DBHF results. Without renormalization the deviations for the saturation density and the binding are even stronger.  
In fig.~\ref{fig:EbindDBHF} the binding energy of the renormalized DDRMF theory  is shown for neutron matter and symmetric nuclear matter. The results are in a fairly good agreement with the DBHF results and much better than the results without renormalization. Although the renormalized isovector coupling functions $g_{\rho}^2$ and $g_{\delta}^2$ show a stronger $\beta$-dependence the assumption of only density dependent couplings is still a good approximation. Extracting those coupling functions at the representative value of $\beta=0.2$ both, the symmetric and the neutron matter EoS, are reproduced with fairly good accuracy.
However, as can be seen from  Table~\ref{tab:sat}, the mapping of DBHF onto the RMF functional leads generally to a shift of the saturation point towards lower densities and a slightly smaller binding energy. This feature which is mainly due to the neglection of the intrinsic momentum dependence of the DBHF self-energy has also been observed in previous works when a similar procedure was applied~\cite{fuchs95,hofmann01}. Although the most realistic, Bonn A leads in DBHF to a slightly too large saturation density~\cite{bm90,gross99} and therefore such a shift of the saturation point is in favor of the DDRMF functional when applied to finite nuclei.
%%%%%%%%%%%%%%%%%%%%%%%%%%%%%%%%%%%%%%%%%%%%%%%%%%%%%%%%%%%%%%%%%%%%%%%%%
\label{sec:rrmft}
\section{Conclusion}
\label{sec:c}
In summary, we present calculations of isospin asymmetric nuclear matter in a relativistic DBHF framework based on projection techniques. The approximation scheme for the treatment of isospin asymmetric nuclear matter has been improved. First of all, the application of the Bonn potential - factually the Bonn A  potential has been used throughout this work - has been modified in order to distinguish between different proton and neutron mass by the evaluation of the potential matrix elements. The modification is essential in the $np$ channel when in-medium matrix elements are evaluated.

Secondly, the $T$-matrix can be represented by a set of six linearly independent Lorentz invariants in the $np$ channel.
This sixth covariant has been chosen as proposed in~\cite{dejong98}. However, in contrast to~\cite{dejong98} we apply still the improved decomposition scheme of the $T$-matrix (subtracted $T$-matrix representation)~\cite{gross99,vandalen04b,vandalen05a,vandalen05b} which minimizes on-shell ambiguities in the determination of the self-energy components. 

It is found that the proton properties are, in particular, sensitive to the consequences of the adaption of the Bonn potential for isospin asymmetric nuclear matter and the introduction of a sixth covariant. The proton optical potential lies a bit  lower and is steeper as compared to ref.~\cite{vandalen05b}, whereas the neutron optical potential is almost unaltered. Furthermore, the neutron effective mass remains practically unaffected, whereas the proton mass experiences a sizable reduction.  The reason is that in neutron rich matter proton properties depend much stronger on contributions from the $np$ channel than neutron properties.

The main properties of isospin asymmetric nuclear matter remain, however, unchanged.
The binding energy shows the expected quadratic dependence on the asymmetry parameter.
Also the depth and the steepness of neutron optical potential 
decreases with increasing asymmetry, whereas the depth and steepness of the proton optical potential
still shows the opposite behavior. A strong density and momentum dependence 
can again be observed for isovector optical  potential. In addition, the isovector optical potential remains almost independent
of the asymmetry parameter $\beta$. Our DBHF calculations
based on projection techniques predict a mass splitting of
$m^*_{D,n} < m^*_{D,p}$ in neutron-rich
matter as expected.
The nonrelativistic mass derived from our DBHF approach still shows the
opposite behavior, which is in agreement with
the results from nonrelativistic BHF calculations
\cite{muether02}.

At present full Brueckner calculations are still too involved for systematic applications in finite nuclei.
However, the density dependent mean field
effective coupling functions, which are obtained from the Brueckner self-energy components,  
parameterize the correlations of the 
$T$-matrix in a handable way. Therefore, these coupling functions can be applied to finite nuclei 
within the framework of DDRMF
theory~\cite{fuchs95}. Doing so, a "naive" parameterization of the DBHF results in terms of a density dependent relativistic mean field functional leads to a poor reproduction of the original EoS. The reason are contributions from Fock-terms which are not present at the mean field level and which have to be incorporated in an effective way. This leads to a renormalization procedure of the coupling functions which absorbs the contributions from the Fock terms.
However, the results with renormalization are in a quite good agreement with the DBHF results. With other words, an accurate reproduction of the DBHF EoS requires a renormalization of the coupling functions.

\newpage
\appendix

\section{Potential matrix elements.}
\label{app:pe}
In this appendix we give the potential matrix elements for scalar, pseudovector, and vector mesons. Overall factors in front are omitted. In contrast to the normal expressions which are used in the Bonn codes~\cite{machleidt01}, we release the assumption of equal nucleon masses. This means that particle one and two can have different masses which leads to additional independent matrix elements. Therefore, one has to consider eight instead of six independent partial wave or helicity matrix elements for scattering of positive energy states. For completeness we present in the following the complete sets of matrix elements for the various OBE amplitudes. We follow the notation of ref.~\cite{machleidt01}.
The potential expressions are presented in terms of helicity states.  First, the expressions for the scalar mesons, the $\sigma$ meson and the $\delta$ meson, are given here
\beqa
<++|V_s^J|++>&=&C_s <++|\phi_s|++>  (I^{(1)}_J+I^{(0)}_J)\\
<++|V_s^J|-->&=&C_s <++|\phi_s|-->  (I^{(1)}_J-I^{(0)}_J)\\
<+-|V_s^J|+->&=&C_s <+-|\phi_s|+->  (I^{(2)}_J+I^{(0)}_J)\\
<+-|V_s^J|-+>&=&C_s <+-|\phi_s|-+>  (I^{(2)}_J-I^{(0)}_J)\\
<++|V_s^J|+->&=&-C_s <++|\phi_s|+-> I^{(3)}_J\\
<++|V_s^J|-+>&=&-C_s <++|\phi_s|+-> I^{(3)}_J\\
<+-|V_s^j|++>&=&-C_s <+-|\phi_s|++> I^{(3)}_J\\
<-+|V_s^J|++>&=&-C_s <+-|\phi_s|++> I^{(3)}_J,
\eeqa
where one has 
\beqa
C_s=\pi g_{NNs}^2
\eeqa
and
\beqa
<\lambda'_1 \lambda'_2|\phi_s|\lambda_1 \lambda_2 > &=& (1-\frac{4 \lambda'_1 \lambda_1 p' p}{\epsilon'_1 \epsilon_1})
(1-\frac{4 \lambda'_2 \lambda_2  p' p}{\epsilon'_2 \epsilon_2})
\eeqa
with $\epsilon_i=E^*_i+m^*_i$. The integrals over the Legendre polynomials $I^{(0)}_J-I^{(6)}_J$ are those given in appendix B of~\cite{machleidt01}.

Secondly, the expression for pseudovector mesons, the $\pi$ meson and the $\eta$ meson, are written as 
\beqa
<++|V_{pv}^J|++>&=&C_{pv} <++|\phi_{pv}|++>  (I^{(1)}_J+I^{(0)}_J)\\
<++|V_{pv}^J|-->&=&C_{pv} <++|\phi_{pv}|-->  (I^{(1)}_J-I^{(0)}_J)\\
<+-|V_{pv}^J|+->&=&C_{pv} <+-|\phi_{pv}|+->  (I^{(2)}_J+I^{(0)}_J)\\
<+-|V_{pv}^J|-+>&=&C_{pv} <+-|\phi_{pv}|-+>  (I^{(2)}_J-I^{(0)}_J)\\
<++|V_{pv}^J|+->&=&-C_{pv} <++|\phi_{pv}|+-> I^{(3)}_J\\
<++|V_{pv}^J|-+>&=&-C_{pv} <++|\phi_{pv}|+-> I^{(3)}_J\\
<+-|V_{pv}^J|++>&=&-C_{pv} <+-|\phi_{pv}|++> I^{(3)}_J\\
<-+|V_{pv}^J|++>&=&-C_{pv} <+-|\phi_{pv}|++> I^{(3)}_J, 
\eeqa
where one has 
\beqa
C_{pv}=\pi \frac{g_{NNpv}^2}{4 M^2}
\eeqa
and
\beqa
<\lambda'_1 \lambda'_2|\phi_{pv}|\lambda_1 \lambda_2> &= &(2\lambda'_1 p'-2 \lambda_1 p) (1+\frac{4 \lambda_1 \lambda'_1 p p'}{\epsilon_1 \epsilon'_1}) \nonumber \\ &&(2\lambda'_2 p'-2 \lambda_2 p) (1+\frac{4 \lambda_2 \lambda'_2 p p'}{\epsilon_2 \epsilon'_2}). 
\label{eq:helpv}
\eeqa
with scaling mass $M$. In eq. (\ref{eq:helpv}) the Blankenbecler sugar or Thomas approximation is used, i.e. the exchanged energy transfer between the two nucleons is restricted to zero. Therefore, the four-momentum transfer is  $(p'-p)^{\mu}=(0,\vecp'-\vecp)$. This approximation is later on also applied for the vector mesons.

Finally, the vector mesons, the $\omega$ meson and the $\rho$ meson,  are treated. The vector-meson exchange potential $V_{v}$ consists of three terms: the vector-vector contribution $V_{vv}$, the tensor-tensor contribution $V_{tt}$, and the mixed vector-tensor contribution $V_{vt}$. The vector-vector part can be written as  
\beqa
<++|V_{vv}^J|++>&=&C_{vv}  [<++|\phi_0|++>  (I^{(1)}_J+I^{(0)}_J) \nonumber \\ &
+ & <++|\phi_{\vecv}|++>  (I^{(1)}_ J - 3 I^{(0)}_J)]\\
<++|V_{vv}^J|-->&=&C_{vv} [<++|\phi_0|-->   (I^{(1)}_J-I^{(0)}_J) \nonumber \\ & + & <++|\phi_{\vecv}|-->   (I^{(1)}_J + 3 I^{(0)}_J)\\
<+-|V_{vv}^J|+->&=&C_{vv} [<+-|\phi_0|+->+<+-|\phi_{\vecv}|+->]   (I^{(2)}_J+I^{(0)}_J)\\
<+-|V_{vv}^J|-+>&=&C_{vv} [<+-|\phi_0|-+>+<+-|\phi_{\vecv}|-+>]   (I^{(2)}_J-I^{(0)}_J)\\
<++|V_{vv}^J|+->&=&-C_{vv} [<++|\phi_0|+->+<++|\phi_{\vecv}|+->]   I^{(3)}_J\\
<++|V_{vv}^J|-+>&=&-C_{vv} [<++|\phi_0|+->+<++|\phi_{\vecv}|+->]   I^{(3)}_J\\
<+-|V_{vv}^J|++>&=&-C_{vv} [<+-|\phi_0|++>+<+-|\phi_{\vecv}|++>]   I^{(3)}_J\\
<-+|V_{vv}^J|++>&=&-C_{vv} [<+-|\phi_0|++>+<+-|\phi_{\vecv}|++>]   I^{(3)}_J,
\eeqa
where one has
\beqa
C_{vv}=\pi g_{NNv}^2,
\eeqa
\beqa
<\lambda'_1 \lambda'_2|\phi_0|\lambda_1 \lambda_2 > &=&(1+\frac{4 \lambda'_1 \lambda_1 p' p}{\epsilon'_1 \epsilon_1})
(1+\frac{4 \lambda'_2 \lambda_2  p' p}{\epsilon'_2 \epsilon_2}),
\eeqa
and
\beqa
<\lambda'_1 \lambda'_2|\phi_{\vecv}|\lambda_1 \lambda_2 > &=&-(\frac{2 \lambda'_1 p'}{\epsilon'_1}+\frac{2 \lambda_1 p}{\epsilon_1})
(\frac{2 \lambda'_2 p'}{\epsilon'_2} +\frac{2 \lambda_2  p}{\epsilon_2}).
\eeqa
The tensor-tensor part is  
\beqa
<++|V_{tt}^J|++>&=&C_{tt} [<++|\phi_{1t}|++>  (I^{(1)}_J+I^{(0)}_J) \nonumber \\&+&
<++|\phi_{1\theta}|++> (I^{(4)}_J+I^{(1)}_J) \label{eq:tt1}
\\ &+& <++|\phi_{\sigma t}|++>  (I^{(1)}_J- 3 I^{(0)}_J)] \nonumber\\
<++|V_{tt}^J|-->&=&C_{tt} [<++|\phi_{1t}|-->  (I^{(1)}_J-I^{(0)}_J) \nonumber \\
&+& <++|\phi_{1\theta}|-->  (I^{(4)}_J-I^{(1)}_J) \\ 
&+& <++|\phi_{\sigma t}|-->  (I^{(1)}_J+3 I^{(0)}_J)]\nonumber \\
<+-|V_{tt}^J|+->&=&C_{tt} [<+-|\phi_{1t}|+->+<+-|\phi_{\sigma t}|+->]  (I^{(2)}_J+I^{(0)}_J) \nonumber \\
&+& <+-|\phi_{1\theta}|+-> (I^{(5)}_J+I^{(1)}_J)  \\
<+-|V_{tt}^J|-+>&=&C_{tt} [<+-|\phi_{1t}|-+>+<+-|\phi_{\sigma t}|-+>]  (I^{(2)}_J-I^{(0)}_J) \nonumber \\
&+&<+-|\phi_{1\theta}|-+>  (I^{(5)}_J-I^{(1)}_J)  \\
<++|V_{tt}^J|+->&=&-C_{tt} [[<++|\phi_{1t}|+->+<++|\phi_{\sigma t}|+->] I^{(3)}_J  \nonumber \\
&+& <++|\phi_{1\theta}|+-> I^{(6)}_J] \\
<++|V_{tt}^J|-+>&=&-C_{tt} [[<++|\phi_{1t}|+->+<++|\phi_{\sigma t}|+->] I^{(3)}_J \nonumber \\
&+& <++|\phi_{1\theta}|+-> I^{(6)}_J] \\
<+-|V_{tt}^J|++>&=&-C_{tt} [[<+-|\phi_{1t}|++>+<+-|\phi_{\sigma t}|++>] I^{(3)}_J \nonumber \\
&+& <+-|\phi_{1\theta}|++> I^{(6)}_J] \\
<-+|V_{tt}^J|++>&=&-C_{tt} [[<+-|\phi_{1t}|++>+<+-|\phi_{\sigma t}|++>] I^{(3)}_J \nonumber \\
&+& <+-|\phi_{1\theta}|++> I^{(6)}_J], \label{eq:tt8}
\eeqa
where one uses 
\beqa
C_{tt}=\pi \frac{g_{NNt}^2}{4 M^2}.
\eeqa 
Furthermore, one has
\beqa
<\lambda'_1 \lambda'_2|\phi_{1\theta}|\lambda_1 \lambda_2> &=&2 p p' (1-\frac{4 \lambda'_1 \lambda_1 p' p}{\epsilon_1 \epsilon'_1})(1-\frac{4 \lambda'_2 \lambda_2 p' p}{\epsilon'_2 \epsilon_2}),
\eeqa
\beqa
<\lambda'_1 \lambda'_2|\phi_{\sigma t}|\lambda_1 \lambda_2> &=& -(\frac{2 \lambda_1 p}{\epsilon_1}+\frac{2 \lambda'_1 p'}{\epsilon'_1})(\frac{2 \lambda_2 p}{\epsilon_2}+\frac{2 \lambda'_2 p'}{\epsilon'_2})(m^*_1+m'^*_1)(m^*_2+m'^*_2)
\nonumber \\ & -& (E'^*_1-E^*_1)(E'^*_2-E^*_2)(\frac{2 \lambda_1 p}{\epsilon_1}-\frac{2 \lambda'_1 p'}{\epsilon'_1}) (\frac{2 \lambda_2 p}{\epsilon_2}-\frac{2 \lambda'_2 p'}{\epsilon'_2}) \nonumber \\
&+&(m^*_2+m'^*_2) (E'^*_1-E^*_1) (\frac{2 \lambda_1 p_1}{\epsilon_1}-\frac{2 \lambda'_1 p'_1}{\epsilon'_1})(\frac{2 \lambda_2 p_2}{\epsilon_2}+\frac{2 \lambda'_2 p'_2}{\epsilon'_2}) \\
&+&(m^*_1+m'^*_1)(E'^*_2-E^*_2)(\frac{2 \lambda_1 p_1}{\epsilon_1}+\frac{2 \lambda'_1 p'_1}{\epsilon'_1})(\frac{2 \lambda_2 p_2}{\epsilon_2}-\frac{2 \lambda'_2 p'_2}{\epsilon'_2}) \nonumber,
\eeqa
and
\beqa
<\lambda'_1 \lambda'_2|\phi_{1t}|\lambda_1 \lambda_2> &=&A_{tt}(1+\frac{4 \lambda'_1 \lambda_1 p' p}{\epsilon_1 \epsilon'_1}) (1+\frac{4 \lambda'_2 \lambda_2 p' p}{\epsilon'_2 \epsilon_2}) \nonumber \\ &+& B_{tt} (1-\frac{4 \lambda'_1 \lambda_1 p' p}{\epsilon_1 \epsilon'_1})(1-\frac{4 \lambda'_2 \lambda_2 p' p}{\epsilon'_2 \epsilon_2}) \nonumber \\ &+& D_{tt} (1+\frac{4 \lambda'_1 \lambda_1 p' p}{\epsilon_1 \epsilon'_1}) (1-\frac{4 \lambda'_2 \lambda_2 p' p}{\epsilon'_2 \epsilon_2}) \nonumber
\\ &+& E_{tt} (1-\frac{4 \lambda'_1 \lambda_1 p' p}{\epsilon_1 \epsilon'_1})(1+\frac{4 \lambda'_2 \lambda_2 p' p}{\epsilon'_2 \epsilon_2})\\ &+& (E'^*_1-E^*_1)(\frac{2 \lambda_1 p}{\epsilon_1}-\frac{2 \lambda'_1 p'}{\epsilon'_1})(2 \lambda'_1 p'+2 \lambda_1 p)  (1-\frac{4 \lambda'_2 \lambda_2 p' p}{\epsilon'_2 \epsilon_2}) \nonumber \\ &+& (E'^*_2-E^*_2)
(\frac{2 \lambda_2 p}{\epsilon_2}-\frac{2 \lambda'_2 p'}{\epsilon'_2})(2 \lambda'_2 p'+2 \lambda_2 p) (1-\frac{4 \lambda'_1 \lambda_1 p' p}{\epsilon'_1 \epsilon_1})        \nonumber
\eeqa
with $A_{tt}=(m^*_1+m'^*_1)(m^*_2+m'^*_2)$, $B_{tt}=(m^*_1+m'^*_1)^2+(m^*_2+m'^*_2)^2+(E'^*_1+E^*_1)(E'^*_2+E^*_2)+p^2 + p'^2$, $D_{tt}=-(E'^*_2+E^*_2+E^*_1+E'^*_1) (m^*_1+m'^*_1)$, and $E_{tt}=-(E'^*_2+E^*_2+E^*_1+E'^*_1) (m^*_2+m'^*_2)$. 
Furthermore, the mixed vector-tensor part is given by
\beqa
<++|V_{vt}^J|++>&=&C_{vt} [<++|\phi_1|++>  (I^{(1)}_J+I^{(0)}_J) \nonumber \\ &+& <++|\phi_{\sigma}|++>  (I^{(1)}_J-3I^{(0)}_J)] \\
<++|V_{vt}^J|-->&=&C_{vt} [<++|\phi_1|-->  (I^{(1)}_J-I^{(0)}_J) \nonumber \\ &+& <++|\phi_{\sigma}|-->  (I^{(1)}_J+3I^{(0)}_J)] \\
<+-|V_{vt}^J|+->&=&C_{vt} [<+-|\phi_1|+->+<+-|\phi_{\sigma}|+->]  (I^{(2)}_J+I^{(0)}_J)\\
<+-|V_{vt}^J|-+>&=&C_{vt} [<+-|\phi_1|-+>+<+-|\phi_{\sigma}|-+>]  (I^{(2)}_J-I^{(0)}_J)\\
<++|V_{vt}^J|+->&=&-C_{vt} [<++|\phi_1|+->+<++|\phi_{\sigma}|+->] I^{(3)}_J\\
<++|V_{vt}^J|-+>&=&-C_{vt} [<++|\phi_1|+->+<++|\phi_{\sigma}|+->] I^{(3)}_J\\
<+-|V_{vt}^J|++>&=&-C_{vt} [<+-|\phi_1|++>+<+-|\phi_{\sigma}|++>] I^{(3)}_J\\
<-+|V_{vt}^J|++>&=&-C_{vt} [<+-|\phi_1|++>+<+-|\phi_{\sigma}|++>] I^{(3)}_J,
\eeqa
where  one has 
\beqa
C_{vt}=\pi \frac{g_{NNv} g_{NNt}}{2 M},
\eeqa
\beqa
<\lambda'_1 \lambda'_2|\phi_1|\lambda_1 \lambda_2> = [A_{vt}+
D_{vt} \frac{16 \lambda'_1 \lambda'_2 \lambda_1 \lambda_2 p'^2 p^2}{\epsilon'_1 \epsilon'_2 \epsilon_1 \epsilon_2}],
\eeqa
and
\beqa
<\lambda'_1 \lambda'_2|\phi_{\sigma}|\lambda_1 \lambda_2> & = & (m^*_1+m'^*_1+m^*_2+m'^*_2) <\lambda'_1 \lambda'_2|\phi_{\vecv}|\lambda_1 \lambda_2> \nonumber \\ &+& [(E'^*_1-E^*_1)(\frac{2 \lambda_1 p}{\epsilon_1}-\frac{2 \lambda'_1 p'}{\epsilon'_1}) (\frac{2 \lambda_2 p}{\epsilon_2}  +\frac{2 \lambda'_2 p'}{\epsilon'_2}) \\ &+& (E'^*_2-E^*_2) (\frac{2 \lambda_1 p}{\epsilon_1}+\frac{2 \lambda'_1 p'}{\epsilon'_1})(\frac{2 \lambda_2 p}{\epsilon_2}-\frac{2 \lambda'_2 p'}{\epsilon'_2}) \nonumber
\eeqa
with $A_{vt}=2(m'^*_1+m'^*_2+m^*_1+m^*_2-E'^*_1-E'^*_2-E^*_1-E^*_2)$ and $D_{vt}=2(\epsilon'_1+\epsilon'_2+\epsilon_1+\epsilon_2)$. For the $\omega$ meson only the vector-vector part contributes, because $g_{NNt}=0$ for the $\omega$ meson. 
\section{Partial wave decomposition}
\label{app:pwd}
For a general two-body reaction with four distinct spin-1/2 particles and ignoring anti-particles,  the number of independent amplitudes is sixteen. Due to parity conservation this number is reduced to eight independent amplitudes.
We denote a helicity amplitude by $<\lambda'_1 \lambda'_2|\phi^J(p',p)|\lambda_1 \lambda_2>$, where $\lambda_i$ and $\lambda'_i$ are the initial and final helicities, respectively. Therefore, we have the following set of amplitudes
\beqa
\phi^J_1(p',p)=<++|\phi^J(p',p)|++> \nonumber\\
\phi^J_2(p',p)=<++|\phi^J(p',p)|--> \nonumber \\
\phi^J_3(p',p)=<+-|\phi^J(p',p)|+-> \nonumber \\
\phi^J_4(p',p)=<+-|\phi^J(p',p)|-+> \\
\phi^J_5(p',p)=<++|\phi^J(p',p)|+-> \nonumber \\
\phi^J_6(p',p)=<++|\phi^J(p',p)|-+> \nonumber \\
\phi^J_7(p',p)=<+-|\phi^J(p',p)|++> \nonumber \\
\phi^J_8(p',p)=<-+|\phi^J(p',p)|++>. \nonumber
\eeqa
To partially decouple this system, it is useful to introduce the following linear combinations of helicity amplitudes
\beqa
T^J_{12,o}=\phi^J_1-\phi^J_2 \nonumber \\
T^J_{34,o}=\phi^J_3-\phi^J_4 \nonumber \\
T^J_{56,o}=\phi^J_5-\phi^J_6 \nonumber \\
T^J_{78,o}=\phi^J_7-\phi^J_8 \\
T^J_{12,e}=\phi^J_1+\phi^J_2 \nonumber \\
T^J_{34,e}=\phi^J_3+\phi^J_4 \nonumber \\
T^J_{56,e}=\phi^J_5+\phi^J_6 \nonumber \\
T^J_{78,e}=\phi^J_7+\phi^J_8. \nonumber 
\eeqa
In solving the coupled scattering equation using the linear combinations of the helicity amplitudes, two subsets of  coupled integral equations,  
\beqa
T^{12}_o= V^{12}_o+\int V^{12}_o T^{12}_o +V^{56}_o T^{78}_o \nonumber \\
T^{34}_o=V^{34}_o+\int V^{34}_o T^{34}_o + V^{78}_o T^{56}_o \label{eq:odd} \\
T^{56}_o=V^{56}_o+\int V^{12}_o T^{56}_o + V^{56}_o T^{34}_o \nonumber \\
T^{78}_o=V^{78}_o+\int V^{78}_o T^{12}_o + V^{34}_o T^{78}_o \nonumber 
\eeqa
and
\beqa
T^{12}_e=V^{12}_e+\int V^{12}_e T^{12}_e + V^{56}_e T^{78}_e \nonumber \\
T^{34}_e=V^{34}_e+\int V^{34}_e T^{34}_e + V^{78}_e T^{56}_e \label{eq:even}\\
T^{56}_e=V^{56}_e+\int V^{12}_e T^{56}_e + V^{56}_e T^{34}_e \nonumber \\
T^{78}_e=V^{78}_e+\int V^{78}_e T^{12}_e + V^{34}_e T^{78}_e, \nonumber
\eeqa
emerge. Equation~(\ref{eq:odd}) is a coupled spin singlet-triplet state, whereas eq.~(\ref{eq:even}) is coupled triplet state. For identical particles the coupled spin singlet-triplet state of eq.~(\ref{eq:odd}) decouples further 
into a decoupled singlet state 
\beqa
T^{12}_o=V^{12}_o+\int V^{12}_o T^{12}_o 
\eeqa
and a decoupled triplet state
\beqa
T^{34}_o=V^{34}_o+\int V^{34}_o T^{34}_o 
\eeqa
due to $^{56}V^J_o=^{78}V^J_o=^{56}T^J_o=^{78}T^J_o=0$.

%%%%%%%%%%%%%%%%%%%%%%%%%%%%%%%%%%%%%%%%%%%%%%%%%%%%%%%%%%%%%%%%%%%%%%%%
%


\begin{thebibliography}{99}
\bibitem{bethe90}
H. A. Bethe, Rev. Mod. Phys. \textbf{62}, 801 (1990). 
%
\bibitem{pethick95}
C. J. Pethick, D. G. Ravenhall, and C. P. Lorentz, Nucl. Phys. \textbf{A584}, 675
(1995). 
%
\bibitem{lattimer91} J. M. Lattimer, C. J. Pethick, M. Prakash, and P. Haensel,
Phys. Rev. Lett. \textbf{66}, 2701 (1991); E. N. E. van Dalen, A. E. L. Dieperink, and J. A. Tjon, Phys. Rev. C \textbf{67}, 065807 (2003).
%
\bibitem{vandalen06b} T. Kl\"ahn, D. Blaschke, S. Typel, E. N. E. van Dalen, A. Faessler, C. Fuchs, T. Gaitanos, H. Grigorian, A. Ho, E. E. Kolomeitsev, M.C. Miller, G. R\"opke, J. Tr\"umper, D. N. Voskresensky, F. Weber, and H. H. Wolter,  Phys. Rev. C \textbf{74}, 035802 (2006). 
%
\bibitem{prakash88}
M. Prakash, T. L. Ainsworth, and J. M. Lattimer, Phys. Rev. Lett. \textbf{61}, 2518 (1988);
L. Engvik, M. Hjorth-Jensen, E. Osnes, G. Bao, and E. Oestgaard,
Phys. Rev. Lett. \textbf{73}, 2650 (1994).
%
\bibitem{zuo04} 
X. R. Zhou, G. F. Burgio, U. Lombardo, H.-J. Schulze, W. Zuo,  
Phys. Rev. C \textbf{69}, 018801 (2004). 
%
\bibitem{tanihata95}
I. Tanihata, Prog. Part. Nucl. Phys. \textbf{35}, 505 (1995);
P. G. Hansen, A. S. Jensen, and B. Jonson,
Annu. Rev. Nucl. Part. Sci. \textbf{45}, 591 (1995).
%
\bibitem{baran05}
V. Baran, M. Colonna, V. Greco, and M. Di Toro, Phys. Rep. \textbf{410}, 335 (2005).
%
\bibitem{reinhard04} 
P.-G. Reinhard, M. Bender,  
Lect. Notes Phys. \textbf{641}, 249 (2004).
% 
\bibitem{rmf} 
P. Ring, Prog. Part. Nucl. Phys. 73, 193 (1996);  
Lect. Notes Phys. \textbf{641}, 175 (2004).
%  
\bibitem{akmal98} 
A. Akmal, V. R. Pandharipande, D. G. Ravenhall,  
Phys. Rev. C \textbf{58}, 1804 (1998).
% 
\bibitem{lejeune00}
A. Lejeune, U. Lombardo, and W. Zuo,  Phys.Lett. B \textbf{477}, 45 (2000).
%
\bibitem{terhaar87} 
B. ter Haar and R. Malfliet, 
Phys. Rep. \textbf{149}, 207 (1987). 
%
\bibitem{bm90} 
R. Brockmann, R. Machleidt, Phys. Rev. C \textbf{42}, 1965 (1990).
%
\bibitem{dejong98}
F. de Jong and H. Lenske,
Phys. Rev. C \textbf{58}, 890 (1998).
%
\bibitem{gross99}
T. Gross-Boelting, C. Fuchs, and Amand Faessler,
Nucl. Phys. \textbf{A648}, 105  (1999).
%
\bibitem{alonso03} D. Alonso and F. Sammarruca, Phys. Rev. C \textbf{67}, 054301 (2003).
%
\bibitem{vandalen04b} E. N. E. van Dalen, C. Fuchs, and Amand Faessler, Nucl. Phys. \textbf{A744}, 227
(2004). 
%
\bibitem{vandalen05a} E. N. E. van Dalen, C. Fuchs, and Amand Faessler, Phys. Rev. Lett. \textbf{95}, 022302 (2005). 
%
\bibitem{vandalen05b} E. N. E. van Dalen, C. Fuchs, and Amand Faessler, Phys. Rev. C \textbf{72}, 065803 (2005). 
%
\bibitem{carlson03}
J. Carlson, J. Morales, V. R. Pandharipande, D. G. Ravenhall, 
Phys. Rev. C \textbf{68}, 025802  (2003); 
W. H. Dickhoff, C. Barbieri, 
Prog. Part. Nucl. Phys. \textbf{52}, 377 (2004). 
% 
\bibitem{gogny} 
J. Decharge, D. Gogny, Phys. Rev. C {\bf 21}, 1568 (1980).
%
\bibitem{fuchs95}
C. Fuchs, H. Lenske, H. H. Wolter, Phys. Rev. C \textbf{52}, 3043 (1995).
%
\bibitem{hofmann01} 
F. Hofmann, C. M. Keil, H. Lenske, Phys.Rev. C \textbf{64}, 034314 (2001). 
%
\bibitem{bombaci91}
I. Bombaci and U. Lombardo, Phys. Rev. C \textbf{44}, 1892 (1991).
%
\bibitem{muether02}
T. Frick, Kh. Gad, H. M\"uther, and P. Czerski, Phys. Rev. C \textbf{65},  034321 (2002); 
W. Zuo, L. G. Cao, B. A. Li, U. Lombardo, and C. W. Shen, Phys.Rev. C \textbf{72}, 014005 (2005).
%
\bibitem{liu02}
B. Liu, V. Greco, V. Baran, M. Colonna, and M. Di Toro, Phys. Rev. C \textbf{65}, 045201 (2002).
%
\bibitem{machleidt89}
R. Machleidt, Adv. Nucl. Phys. \textbf{19}, 189 (1989).
%
\bibitem{horowitz87}
C. J. Horowitz and B. D. Serot,
Nucl. Phys. \textbf{A464}, 613 (1987).
%
\bibitem{sehn97}
L. Sehn, C. Fuchs, and A. Faessler,
Phys. Rev. C \textbf{56}, 216  (1997).
%
\bibitem{bethe63}
H. A. Bethe, B. H. Brandow, and A. G. Petschek,
Phys. Rev. \textbf{129}, 225 (1963).
%
\bibitem{erkelenz74}
K. Erkelenz,
Phys. Rep. C \textbf{13}, 191 (1974).
%
\bibitem{haftel70}
M. I. Haftel and F. Tabakin,
Nucl. Phys. \textbf{A158}, 1 (1970).
%
\bibitem{rose57}
M. Rose, Elementary Theory of Angular Momentum (Wiley, New York, 1957).
%
\bibitem{tjon85a}
J. A. Tjon and S. J. Wallace,
Phys. Rev. C \textbf{32}, 267 (1985).
%
\bibitem{wiringa95} 
R. B. Wiringa, V. G. J. Stoks, and R. Schiavilla, Phys. Rev. C \textbf{51}, 38 (1995). 
%
\bibitem{pudliner95}
B. S. Pudliner, V. R. Pandharipande, J. Carlson, and R. B. Wiringa, Phys. Rev. Lett. \textbf{74}, 4396 (1995).
%
\bibitem{qmc} 
S. Gandolfi, F. Pederiva, S. Fantoni, and K. E. Schmidt, nucl-th/0607022.
%
\bibitem{stoicea04}
G. Stoicea {\it et al.} [FOPI Coll.], 
Phys. Rev. Lett. {\bf 92}, 072303 (2004).
%
\bibitem{sturm01} 
C. Sturm {\it et al.} [KaoS Coll.],  
Phys. Rev. Lett. {\bf 86}, 39 (2001); C. Fuchs, A. Faessler, E. Zabrodin, Y.M. Zheng,   
Phys. Rev. Lett. {\bf 86}, 1974 (2001); C. Fuchs, Prog. Part. Nucl. Phys. {\bf 56}, 1 (2006).
%
\bibitem{nice05}
D. J. Nice, E. M. Splaver, I. H. Stairs, O. L\"ohmer, A. Jessner, M. Kramer, and J.M. Cordes, Astrophys. J. \textbf{634}, 1242 (2005).
%
\bibitem{li04}
B.-A. Li, Phys.Rev. C \textbf{69},  064602  (2004).
%
\bibitem{machleidt01}
R. Machleidt, Phys.Rev. C \textbf{63}, 024001 (2001). 
%
\end{thebibliography}
\end{document}